\documentclass[11pt,a4paper]{article}
\pdfoutput=1

\usepackage{bm}
\usepackage{amsfonts}
\usepackage[dvipdfmx]{graphicx}
\usepackage{jcappub}

\title{\boldmath Neutrino halo effect on collective neutrino oscillation in iron core-collapse supernova model of a 9.6 $M_{\odot}$ star}

\author[a]{Masamichi Zaizen,}

\author[b]{John F. Cherry,}
\author[c]{Tomoya Takiwaki,}
\author[d]{Shunsaku Horiuchi,}
\author[e,f]{Kei Kotake,}
\author[a]{Hideyuki Umeda,}
\author[a]{and Takashi Yoshida}

\affiliation[a]{Department of Astronomy, Graduate School of Science, University of Tokyo, Tokyo 113-0033, Japan}
\affiliation[b]{Department of Physics, University of South Dakota, Vermillion, SD 57069, USA}
\affiliation[c]{Division of Science, National Astronomical Observatory of Japan, 2-21-1 Osawa, Mitaka, Tokyo 181-8588, Japan}
\affiliation[d]{Center for Neutrino Physics, Department of Physics, Virginia Tech, Blacksburg, VA 24061, USA}
\affiliation[e]{Department of Applied Physics, Fukuoka University, Nanakuma 8-19-1, Johnan, Fukuoka 814-0180, Japan}
\affiliation[f]{Research Insitute of Stellar Explosive Phenomena, Fukuoka University, Nanakuma 8-19-1, Johnan, Fukuoka 814-0180, Japan}

\emailAdd{mzaizen@astron.s.u-tokyo.ac.jp}
\emailAdd{jsquair@gmail.com}
\emailAdd{takiwaki.tomoya@nao.ac.jp}
\emailAdd{horiuchi@vt.edu}
\emailAdd{kkotake@fukuoka-u.ac.jp}
\emailAdd{umeda@astron.s.u-tokyo.ac.jp}
\emailAdd{tyoshida@astron.s.u-tokyo.ac.jp}

\date{\today}

\abstract{
We extend the multi-angle computational framework and investigate the time evolution of the neutrino halo on collective neutrino oscillation in the core collapse of an iron core progenitor. We find that in the case of the $9.6\, \rm M_\odot$ progenitor adopted in this work, there are windows of time when the effects of neutrino halo and collective neutrino oscillation are not simultaneously large. 
Inside the shock, the impact of the inward-scattered halo neutrino cannot in general be neglected compared to the outward-propagating neutrino flux. However, during early epochs, collective neutrino oscillation is effectively shut down by multi-angle matter suppression. During the intermediate epoch, collective neutrino oscillation is not suppressed, but its onset radius is beyond the still relatively small explosion shock front where the halo is prominent. We also find in the case of the $9.6\, \rm M_\odot$ progenitor the halo neutrinos induce a delay in the onset of collective neutrino oscillations. This causes novel flavor conversions which sharpen collective neutrino oscillation spectral features. We predict that the inclusion of neutrino halo effects makes neutrino signals that are more clearly distinct from thermal emission that when halo neutrinos are omitted. 
}

\begin{document}
\maketitle


\section{Introduction}
Neutrinos are intensively emitted from supernovae and have the potential to carry with them information on the central properties of the explosion \cite{Langanke:2003a, Mezzacappa:2005a, Woosley:2005a, Kotake:2006a, Janka:2012a, Janka:2017a, Burrows:2013a, Foglizzo:2015a}.
Kamiokande-II and Irvine-Michigan-Brookhaven detector detected eleven and eight neutrinos, respectively, from SN1987A which emerged in the Large Magellanic Cloud \cite{Hirata:1987a, Bionta:1987a}.
This observation indicated the importance of neutrinos in core-collapse supernovae (CCSNe) \cite{Vissani:2015a}.
However, there are still many open questions in supernova physics.
We do not completely understand what happens in the core of exploding stars.
Central densities are sufficient to trap even neutrinos, which then diffuse and are emitted from the proto-neutron star at the core.  The energy spectra for different lepton flavors of neutrinos depend sensitively on the equation of state of nuclear matter in the emission region.
More precise models of neutrino emission and flavor transformation enable us to enrich our understanding of CCSNe.
Current and future neutrino detectors expect that several thousands of neutrino events may be observed from a galactic signal (e.g., \cite{Scholberg:2012a}).
In this case, we will obtain detailed time evolution of the neutrino spectra.
This information is expected to reveal many properties inside CCSNe and, hopefully, enable us to test our current understanding of the yet-uncertain supernova physics (e.g., \cite{Horiuchi:2018b}).

However, we can not get original neutrino spectra directly from observation without accounting for neutrino oscillations.
Neutrinos from CCSNe undergo flavor conversions and arrive at Earth as mixed states \cite{Duan:2010a, Mirizzi:2016a, Chakraborty:2016c}. 
First, neutrino oscillation in vacuum is a $SU(3)$ oscillator system typically parameterized by two mass-squared differences, three mixing angles, and a complex phase.
With the exception of the complex phase, these parameters are currently well-known thanks to the hard work of many neutrino experiments (see summary in Ref.~\cite{PDG18}).
Furthermore, neutrinos undergo flavor conversion with background matter described by the Mikheyev-Smirnov-Wolfenstein (MSW) effect which occurs at two typical electron densities \cite{MSW78, MSW85}.
These two types of neutrino oscillation are linear effects and can easily be grasped in solar and atmospheric neutrino probes.
On the other hand, a third phenomenon becomes potentially dominant in CCSNe which is triggered by neutrino-neutrino coherent forward scattering interactions in the high neutrino flux regions near the proto-neutron star.
Known as collective neutrino oscillation, neutrinos propagating on intersecting trajectories develop quantum coherence in their flavor oscillation with adjacent neutrinos.  While this process has been studied intensively for more than a decade, the non-linear nature of the phenomenon makes the outcome of flavor oscillation impossible to predict without computational modeling.
One of the known outcomes of collective neutrino oscillation is a peculiar behavior, called a spectral split or swap, where the neutrino spectra exhibit flavor exchange only above a critical energy; this behavior is not seen in linear effects \cite{Fogli:2007a}.

The neutrino-neutrino coherent forward scattering interaction depends strongly on the relative trajectories of intersecting neutrinos.
Treatment of collective neutrino oscillation often employs the multi-angle approximation, which considers polar angular distributions of emitted neutrinos in detail.
Neutrinos along different trajectories are forced to synchronize into coherent oscillation modes by the self-interactions and maintain coherence among different angular paths.
On the other hand, this trajectory dependence enhances matter-induced decoherence.
Background electrons cause the phase dispersion due to the travel distance difference.
This phenomenon is called multi-angle matter suppression, and it weakens collective neutrino oscillation ~\cite{Esteban:2008b}.
It does not occur under the single-angle approximation which ignores angular dependence of neutrino wave function.
The competition between the neutrino self-interaction and the matter-induced phase dispersion is important under realistic supernova environments \cite{Chakraborty:2011a, Chakraborty:2011b, Wu:2015a, Zaizen:2018a}.

These interesting features are based on the ``bulb model'', which imposes many assumptions on neutrino emissions and background environments.
Recent studies have revealed that symmetry breaking can enhance new behaviors of collective neutrino oscillation \cite{Cherry:2012b, Cherry:2013a, Cirigliano:2018a,Sarikas:2012b, Raffelt:2013a, Mirizzi:2013a, Chakraborty:2014b, Duan:2015a, Dasgupta:2015a, Chakraborty:2016a, Chakraborty:2016b, Dasgupta:2017a, Abbar:2018a, Delfan:2019a}.
Some of these instabilities can overcome the multi-angle matter suppression and may have large influence on observed neutrino spectra.
However, few detailed calculations using realistic supernova models have been performed due to the numerical complexity.

Many groups have extended the bulb model and tackled collective neutrino oscillation.
For example, Cherry et al. suggested that some of the neutrinos experience a direction-changing scattering with nucleon/nucleus outside the neutrino sphere \cite{Cherry:2012b, Cherry:2013a}.
These scattered neutrinos produce a \lq\lq halo\rq\rq\ flux of neutrinos with large intersection angles with outgoing neutrinos and may have a large impact on the neutrino-neutrino interaction.
Especially, the authors pointed out that inwardly scattered neutrinos can destroy the bulb framework and produce numerical obstacles under some circumstances.
The inward-going flux depends on the background matter density and composition, integrated over the interior of the entire CCSN envelope.
The previous work employed an electron capture CCSN model whose density profile steeply drops at $r > 1000\mathrm{~km}$.
Thus, the inward halo flux decreases outside the steep density gradient, and collective neutrino oscillation with bulb+halo model was able to be performed safely.
Ref.~\cite{Cirigliano:2018a} investigated including the inward-scattered neutrinos self-consistently within simplified collective neutrino oscillation calculations.
The authors showed that main features in collective neutrino oscillation with single-angle approximation, which averages neutrino states with different angular modes, are maintained under the inclusion of the halo flux by searching for a relaxation solution under the assumption of a static total neutrino flux.
Ref.~\cite{Sarikas:2012b} studied the impact of the halo effects on the multi-angle matter suppression with a $15 M_\odot$ progenitor.
Stability analysis including broader neutrino angular distributions showed that the matter-induced phase dispersion dominates the neutrino self-interaction even in the presence of halo neutrinos during the accretion phase of CCSNe with dense envelopes.
The authors suggested that the halo effects do not change the complete suppression due to high matter density.
However, the potential impacts of the neutrino halo during other epochs and other progenitors have not yet been investigated.

In this paper we perform the first ever numerical study of collective neutrino oscillation in the core collapse of a $9.6M_\odot$ iron core progenitor including a consideration of the halo flux in a multi-angle solution framework. 
For completeness, we compare results of our full calculation (\lq with halo\rq ) to a calculation which omits halo neutrino scattering (\lq no halo\rq ) and also to a calculation which omits collective neutrino oscillation entirely (\lq no collective neutrino oscillation\rq ). We then evaluate the time evolution of the signal and the event rates observed by neutrino detectors.  Our principle finding is that, for our chosen progenitor, the presence of the halo neutrinos enhances the detectability of collective neutrino oscillation spectral features.

In section \ref{sec:2}, we introduce our calculation method and employed supernova model.
We present the results of halo distribution, flavor conversions, and detectability in section \ref{sec:3}.
Finally we summarize the conclusions of our study in section \ref{sec:4}.

\section{Numerical setup}
\label{sec:2}

\subsection{Supernova model} 
We perform a two-dimensional (axi-symmetric) CCSN simulation with a $9.6\, \rm M_{\odot}$ zero-metallicity model (Z9.6) provided by A. Heger (2017, private communication, this model is a extension of Heger et al. 2010 \cite{Heger:2010a} toward the lower mass).
This progenitor is a non-rotating star and has an iron-core in the center, different from O-Ne-Mg progenitor used in previous work \cite{Cherry:2013a}.

\begin{figure}[h]
	\centering
    \includegraphics[width=0.6\linewidth]{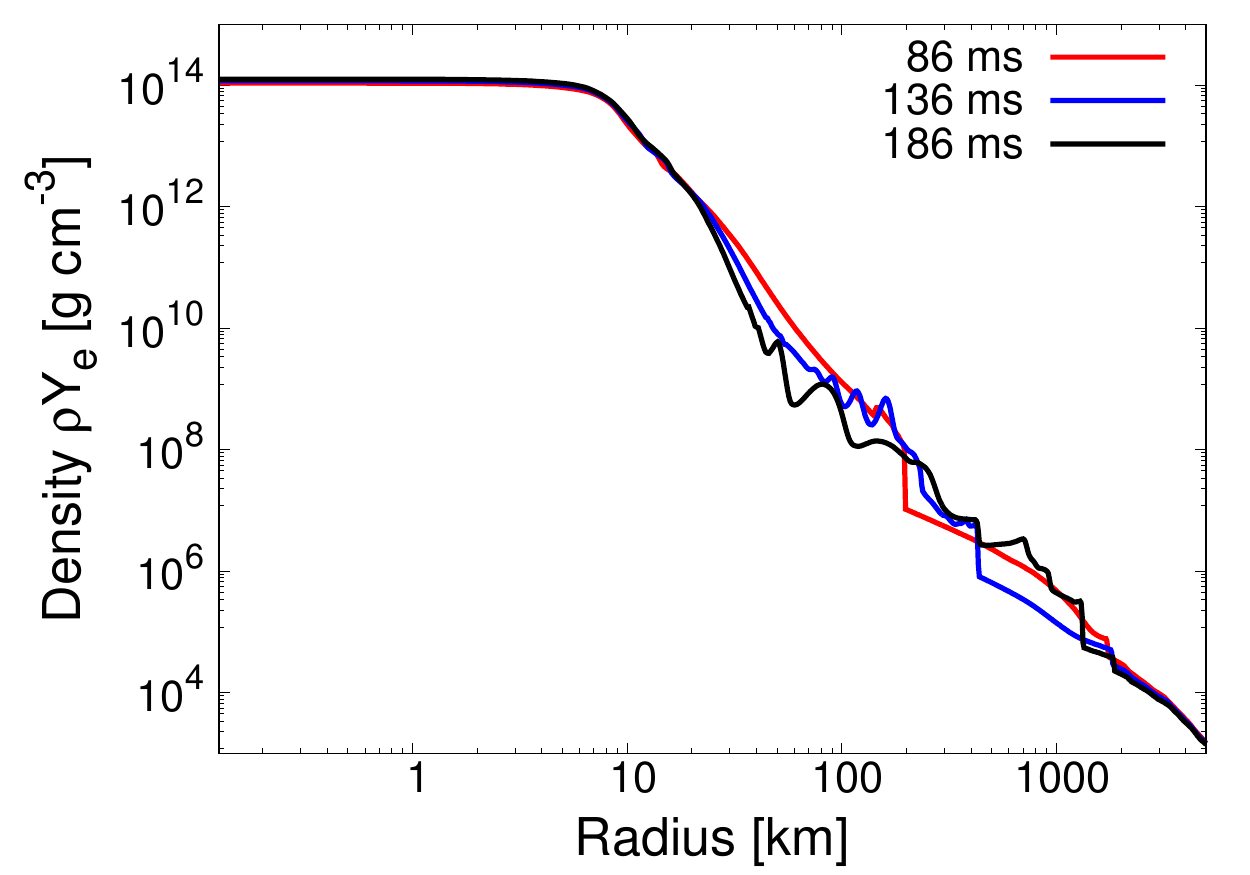}
	\caption{Electron density profile along the north pole at postbounce time $t_{\rm pb} = 86, 136$, and $186\mathrm{~ms}$.
	Shock wave propagates from $200\mathrm{~km}$ to $1000\mathrm{~km}$.
	}
	\label{fig:north_density}
\end{figure}

\begin{figure*}[htbp]
	\begin{minipage}{0.5\hsize}
	    \centering
	    \includegraphics[width=1.0\linewidth]{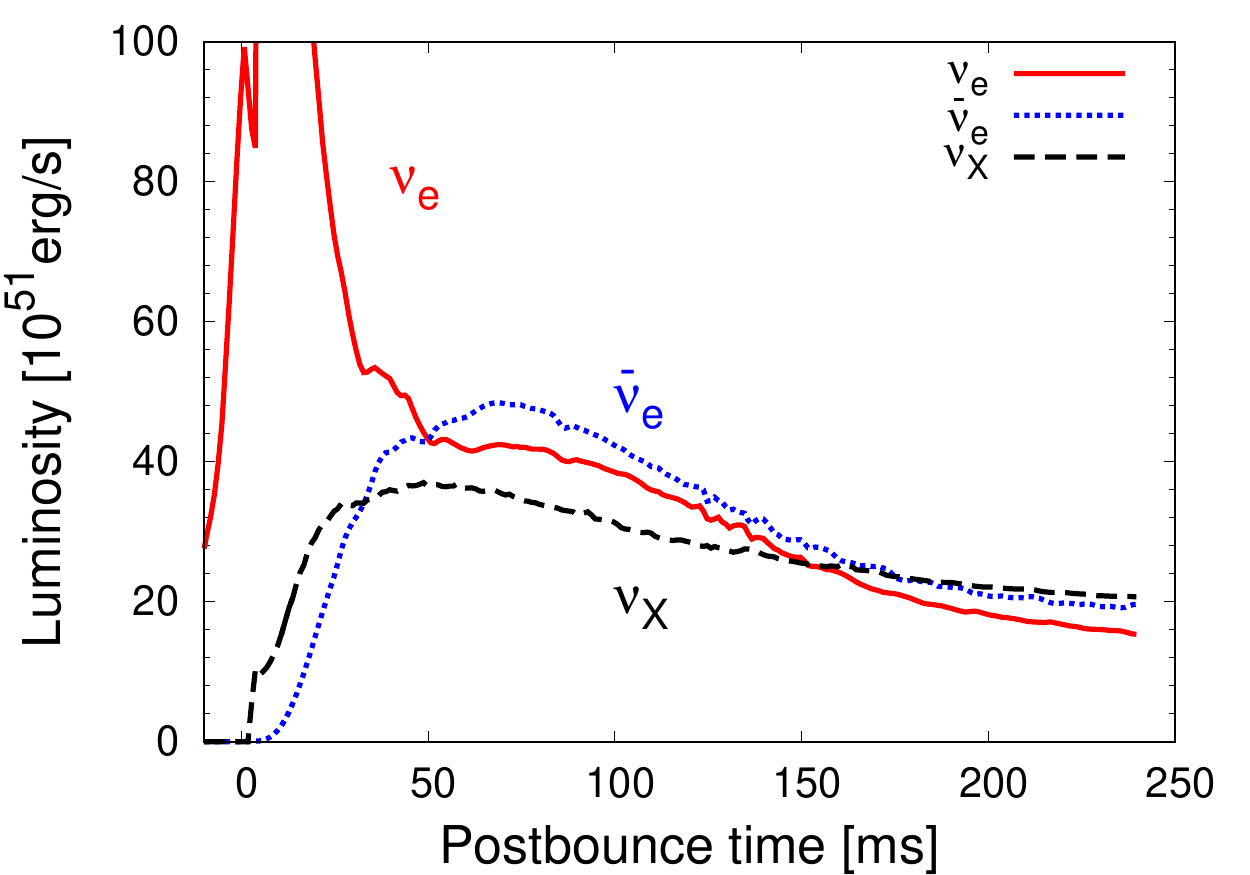}
	\end{minipage}
	\begin{minipage}{0.5\hsize}
	    \centering
    	\includegraphics[width=1.0\linewidth]{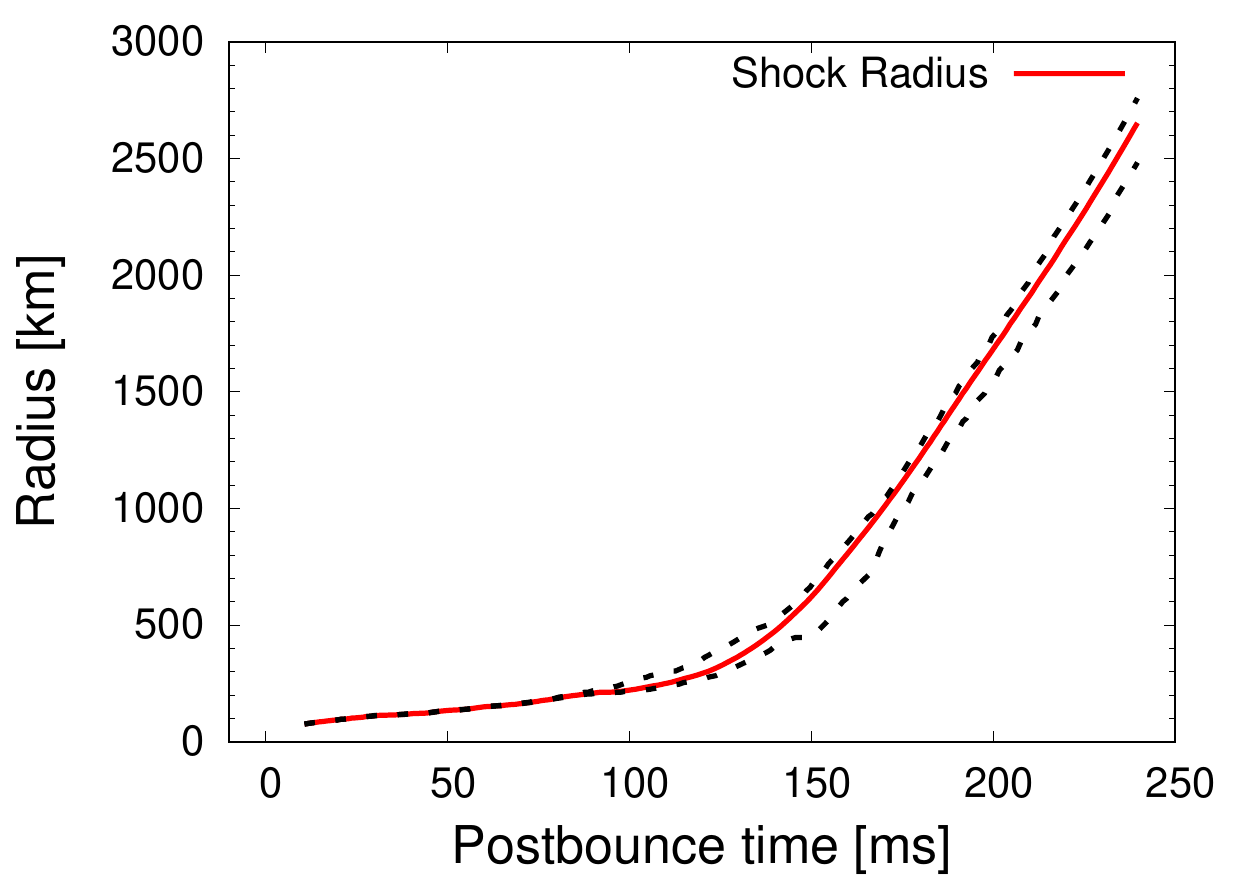}
	\end{minipage}
	
	\begin{minipage}{0.5\hsize}
	    \centering
    	\includegraphics[width=1.0\linewidth]{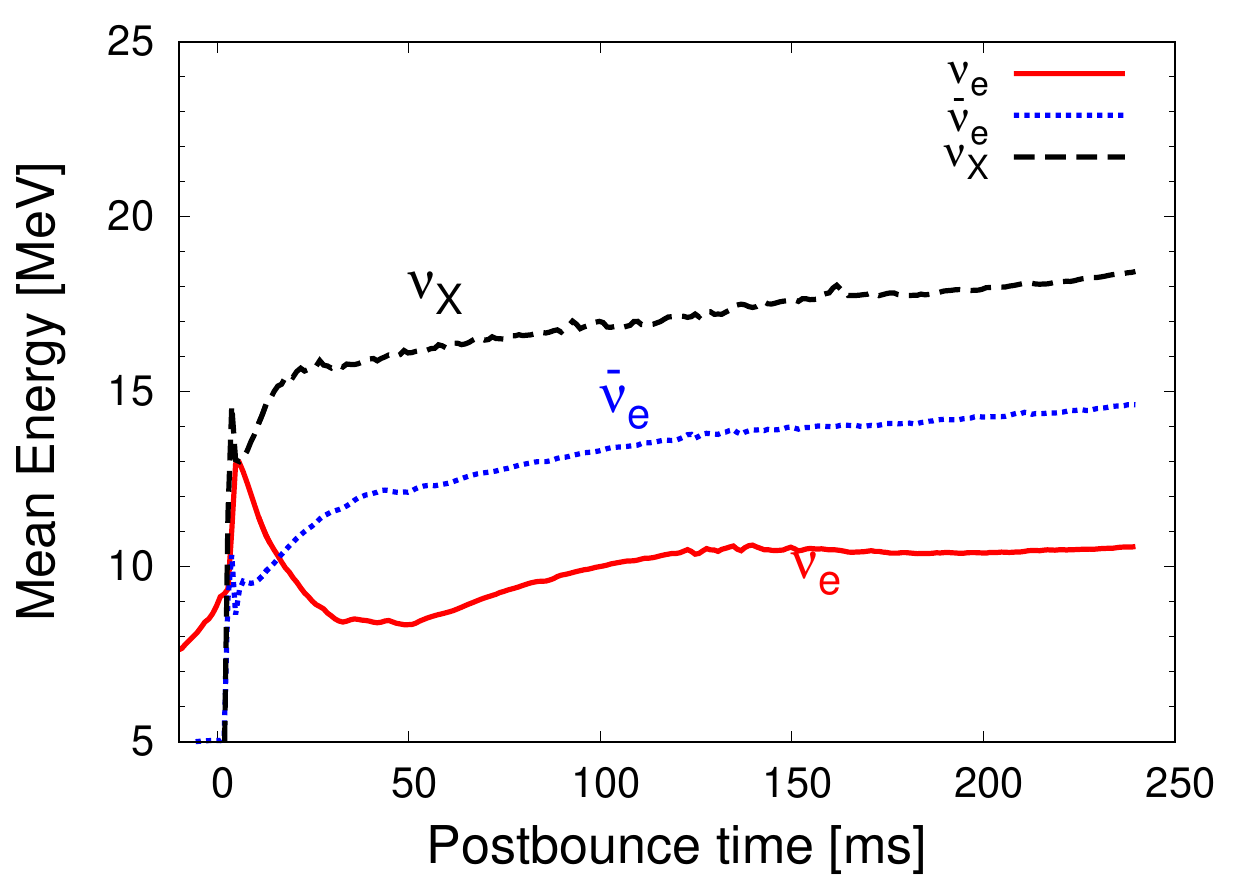}
	\end{minipage}
	\begin{minipage}{0.5\hsize}
	    \centering
    	\includegraphics[width=1.0\linewidth]{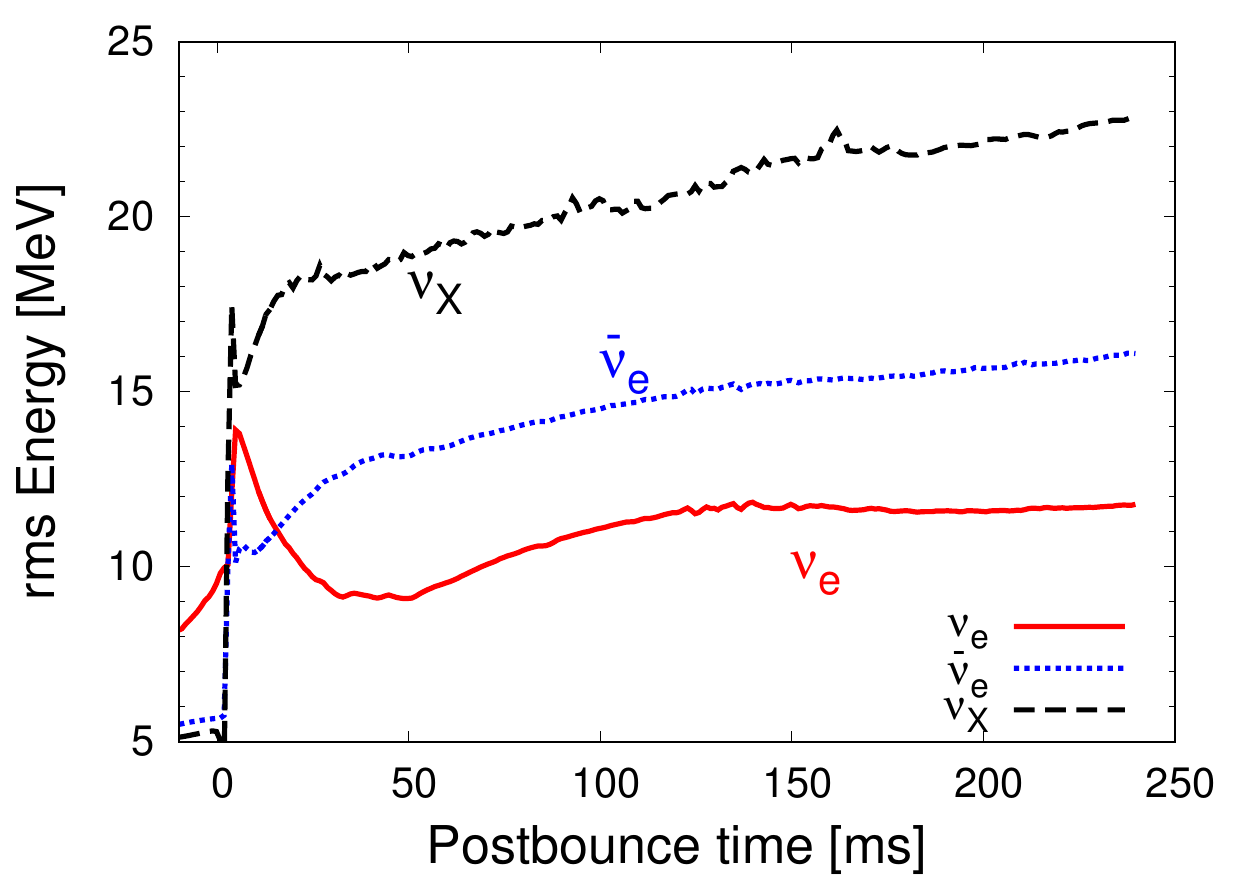}
	\end{minipage}
	\caption{The top panels show the time evolution of neutrino luminosity (left) and shock wave radius (right).
	In the bottom panels, we show averaged energy (left) and rms energy (right).
	In neutrino luminosity and average energy, $\nu_e, \bar{\nu_e}$, and $\nu_X$ are red solid, blue dotted, and black dashed line, respectively.
	For the shock radius figure, dotted lines show the maximum and minimum radius.
	}
	\label{fig:SN_model}
\end{figure*}

The hydrodynamical simulation is performed by 3DnSNe code (recent application reference \cite{Takiwaki:2018a, OConner:2018a}), and we show a selection of results for the electron density profile along the north polar direction at $86\,\rm ms,\ 136\,\rm ms$, and $186\,\mathrm{ms}$ in Figure \ref{fig:north_density}.
This two-dimensional simulation is computed on a spherical polar coordinate with spatial resolution of $(N_r, N_{\Theta})=(512,128)$.
This radial grid covers from the center to an outer boundary of $5000\mathrm{~km}$.
A piecewise linear method with geometrical correction is used to reconstruct variables at the cell edge, where a modified van Leer limiter is employed to satisfy the condition of total variation diminishing (TVD) \cite{Mignone:TVD}.
The numerical flux is calculated by HLLC solver \cite{Toro:HLLsolver}.
We adopt the equation of state by Lattimer \& Swesty with incompressibility of $K=220\mathrm{~MeV}$ \cite{LS91}. 
These features of the time evolution are related to the shock propagation.
Numerical explosion simulations under spherical symmetry are apt to fail the shock revival and do not provide the correct neutrino signals. 
In order to investigate the time evolution of halo effects and collective neutrino oscillation, successfully exploding supernova models are required. 
We have found the Z9.6 model explodes successfully even in spherically symmetric simulation (consistent with \cite{Melson:2015a}), but we here employ a two-dimensional simulation to understand more general neutrino halo structure.
The halo structure strongly couples to the hydrodynamics of supernovae and two-dimensional halo effects are different from their spherical symmetric counterparts \cite{Delfan:2020a,Abbar:2020a,Glas:2020a}. 
We finally calculate collective neutrino oscillations along the north polar direction from the accretion phase until the shock revival.

Figure \ref{fig:SN_model} represents the time evolution of neutrino luminosities $L_{\nu}$, averaged neutrino energies $\langle E_{\nu}\rangle$, rms energy $\sqrt{\langle E_{\nu}^2\rangle}$, and shock radius.
In these neutrino properties, $\nu_X$ means nonelectron type neutrinos $\nu_{\mu}, \bar{\nu}_{\mu}, \nu_{\tau}$, and $\bar{\nu}_{\tau}$.
We approximate neutrino spectra on the surface of neutrino sphere by a gamma distribution \cite{Keil:2003a, Tamborra:2012b, Tamborra:2014a}:
\begin{eqnarray}
f(E_{\nu}) = \frac{(E_{\nu})^{\xi}}{\Gamma_{\xi+1}}\left(\frac{\xi+1}{\langle E_{\nu}\rangle}\right)^{\xi+1}\exp\left[-\frac{(\xi+1)E_{\nu}}{\langle E_{\nu}\rangle}\right],
\end{eqnarray}
where $\Gamma_{\xi+1}$ is the Gamma function.
This $\xi$ is a pinching parameter given by
\begin{eqnarray}
\xi = \frac{\langle E_{\nu}^2\rangle-2\langle E_{\nu}\rangle^2}{\langle E_{\nu}\rangle^2 -\langle E_{\nu}^2\rangle}.
\end{eqnarray}
This shock radius shows the evolution of multi-dimensionality.
The radius of shock wave is almost spherically symmetric before $t_{\mathrm{pb}} \sim 100\mathrm{~ms}$ (with $t_{\mathrm{pb}}$ the postbounce time) while two dimensionality evolves and difference between maximum and minimum shock radius emerges after $t_{\mathrm{pb}} \sim 100\mathrm{~ms}$.

\subsection{Neutrino halo}
\label{sec:2_halo}
Inclusion of the neutrino halo effect within the framework of collective neutrino oscillation calculations has received a renewed burst of attention \cite{Cirigliano:2018a,Richers:2019a} due, in part, to the recent direct detection of coherent enhancement of elastic neutrino-nucleus scattering by the COHERENT experiment \cite{COHERENT:2017a}.  The COHERENT result has placed the presence of the neutrino halo within CCSNe on firmer theoretical ground than collective neutrino oscillation itself, as $\nu-\nu$ coherent forward scattering has not yet been directly observed.  Nevertheless, the problem of including the neutrino halo, itself, within the collective neutrino oscillation framework remains a difficult task of selecting a set of reasonable approximations which allow for the solution of the neutrino flavor transformation equations of motion (EOM) with available computing power.  Our process for generating collective neutrino oscillation predictions from CCSNe simulations follows 4 steps:
\begin{enumerate}
\item Post processing single time snapshots of the neutrino emission from the hydrodynamical simulation of the 3DnSNe code to account for the re-direction of halo neutrinos and creating a 4D map of the energy and angular distribution of halo neutrinos in all radial and angular zones.
\item For each trajectory in each time snapshot, we determine the radii where collective neutrino oscillation is suppressed by multi-angle matter interactions.  We take a point slightly inside the radius where multi-angle suppression ceases as a starting point for the full collective neutrino oscillation calculation.
\item To safely proceed with a collective neutrino oscillation calculation, we verify that the contribution from the halo neutrinos to the neutrino flavor transformation EOM at all radii above the starting point identified in step 2 is suitably ``small'' where this is set to be less than $10\%$ of the $\nu - \nu $ forward scattering interaction.
\item Use the map of the halo neutrinos generated in the first step to populate the outward directed neutrino emission trajectory bins (including the bulb emission) at the calculation starting point and performing a halosphere style~\cite{Cherry:2013a} collective neutrino oscillation calculation. In this sense, we are creating a validation check for what regions of the envelope are safe to treat in terms of the initial condition formulation of the collective neutrino oscillation problem, and what regions require the treatment of the full boundary value problem.  We have found that for the example of the Z9.6 progenitor, we can safely employ the initial condition solution method we describe for the epochs of the SNe explosion which exhibit collective neutrino oscillations.
\end{enumerate}

Our method relies heavily on the results of the first step above. We therefore first explain in detail how the halo neutrino population is calculated before outlining the remaining steps.  In order to generate a map of the halo neutrino population, we calculate the single-scattering contribution for the zero energy transfer neutrino-nucleus interaction cross section to the neutrino transport processes which have already been solved by the 3DnSNe code.  This scattering process is typically included in CCSNe simulations because of its contribution to neutrino trapping during core collapse and because the {\it non-zero} energy transfer portion of the cross section contributes to neutrino energy deposition.  However, the zero energy transfer interaction scatters neutrinos at wide angles, transporting neutrinos along non-radially directed trajectories, is explicitly omitted in the ray-by-ray neutrino transport approximation.  For this reason, we must solve a simplified set of Boltzmann transport equations to calculate the halo neutrino population in a post-hoc fashion, creating a map of the wide angle scattering out of and into each zone in the hydrodynamic simulation, labeled with radial coordinate $r_i$ and polar angle coordinate $\Theta_j$.  

First calculated by Tubbs and Schramm~\cite{Tubbs:1975a}, the enhanced neutrino-nucleus interaction cross section for a nucleus with total nucleons, $A$, proton number, $Z$, and neutron number, $N$, is to leading order,
\begin{equation}
\sigma \left[ E_\nu , \left( Z,N \right)\right] \approx \frac{G_F^2}{\pi} E_{\nu}^2 \left[ \frac{1}{2}\left( C_A - C_V )\right) A + \frac{1}{2} \left( 2 - C_A - C_V ) \right) \left( Z-N \right) \right]^2\, ,
\label{TnS}
\end{equation}
with the Fermi coupling constant, $G_{\mathrm{F}}$, the incoming neutrino energy, $E_\nu$, and weak interaction coupling constants $C_A = 1/2$ and $C_V = 1/2 + 2\sin^2\theta_W = 0.9446$, taking the Weinberg angle to be $\sin^2 \theta_W = 0.2223$.  This cross section is approximately isotropic and thus responsible for redirecting a portion of the neutrino flux emitted during a CCSN along all trajectories within the envelope of the exploding star.

With the coherently enhanced cross section we are able to solve a greatly simplified set of Boltzmann transport equations for neutrinos emitted along each radial ray of the simulation.  We choose to remain consistent with the neutrino transport scheme of the original CCSNe simulation, solving neutrino scattering into the halo population in the ray-by-ray approximation as our initial step,
\begin{equation}
\frac{d }{dr} \rho^{\rm halo}\left( r,\Theta_j, \alpha, E_\nu\right) = \Gamma^{\rm halo}\left( r,\Theta_j, E_\nu \right) \rho^{\rm ray} \left( r,\Theta_j, \alpha , E_\nu \right) \, ,
\label{source}
\end{equation}
and,
\begin{equation}
\frac{d }{dr} \rho^{\rm ray}\left( r,\Theta_j, \alpha, E_\nu\right) = -\Gamma^{\rm halo} \left( r,\Theta_j, E_\nu \right) \rho^{\rm ray} \left( r,\Theta_j, \alpha , E_\nu \right) \, ,
\label{sink}
\end{equation}
with the neutrino scattering rate,
\begin{equation}
\Gamma^{\rm halo} \left( r,\Theta_j, E_\nu \right) = \sum_{\beta} \sigma \left[ E_\nu , \left( Z,N \right)_{\beta}\right] \rho_{\beta}\left( r,\Theta_j \right)\, ,
\end{equation}
taking $\Theta_j$ and the index $j$ to denote the trajectory and angular bin, respectively, of the 2D CCSNe data set, $\alpha$ indexes the neutrino flavor state, $\beta$ indexes the species of nucleus/nucleon for which we calculate scattering rates, and the local density of a nuclear species along a given ray is $\rho_{\beta}\left( r,\Theta_j \right)$.  For simplicity, the index $\beta$ runs over four distinct species: protons, neutrons, ${^4}\mathrm{He}$, and a \lq\lq heavy\rq\rq\ species given by $\langle \left( Z,N\right)\rangle$ for all nuclei heavier than He.  

The population of halo neutrinos which are sourced by scattering in each radial ($i$) and angular ($j$) zone, $ f^{\rm halo}_{i,j}\left( \alpha, E_\nu\right)$, of the CCSNe simulation are then calculated by integrating Equations~(\ref{source}) and ~(\ref{sink}) outward along each ray for the simulation polar coordinate $\Theta_j$, starting at the neutrino sphere, for each radial zone.  Note that solving Equations (4) and (5) for each ray is not the final solution for the halo, but is instead the distribution of scattered neutrino sources which contribute to the final halo distribution at all points $\{r_i,\Theta_j\}$ within the original SNe simulation.  The magnitude of the population of halo neutrinos created by coherently enhanced scattering is $\propto A$, the nuclear mass number of nuclei within the envelope, which is in turn due to the $\propto A^2$ scaling of Equation~(\ref{TnS}) combined with the $\propto A^{-1}$ scaling of the density of atomic nuclei.  For CCSNe which arise from iron-core progenitor stars (estimated to be $\sim 70\%$ of the galactic population \cite{adams:2013a}), the magnitude of the halo effect is proportionately larger than the previously studied O-Ne-Mg core collapse case.  

\begin{figure}[h]
	\centering
    \includegraphics[width=0.6\linewidth]{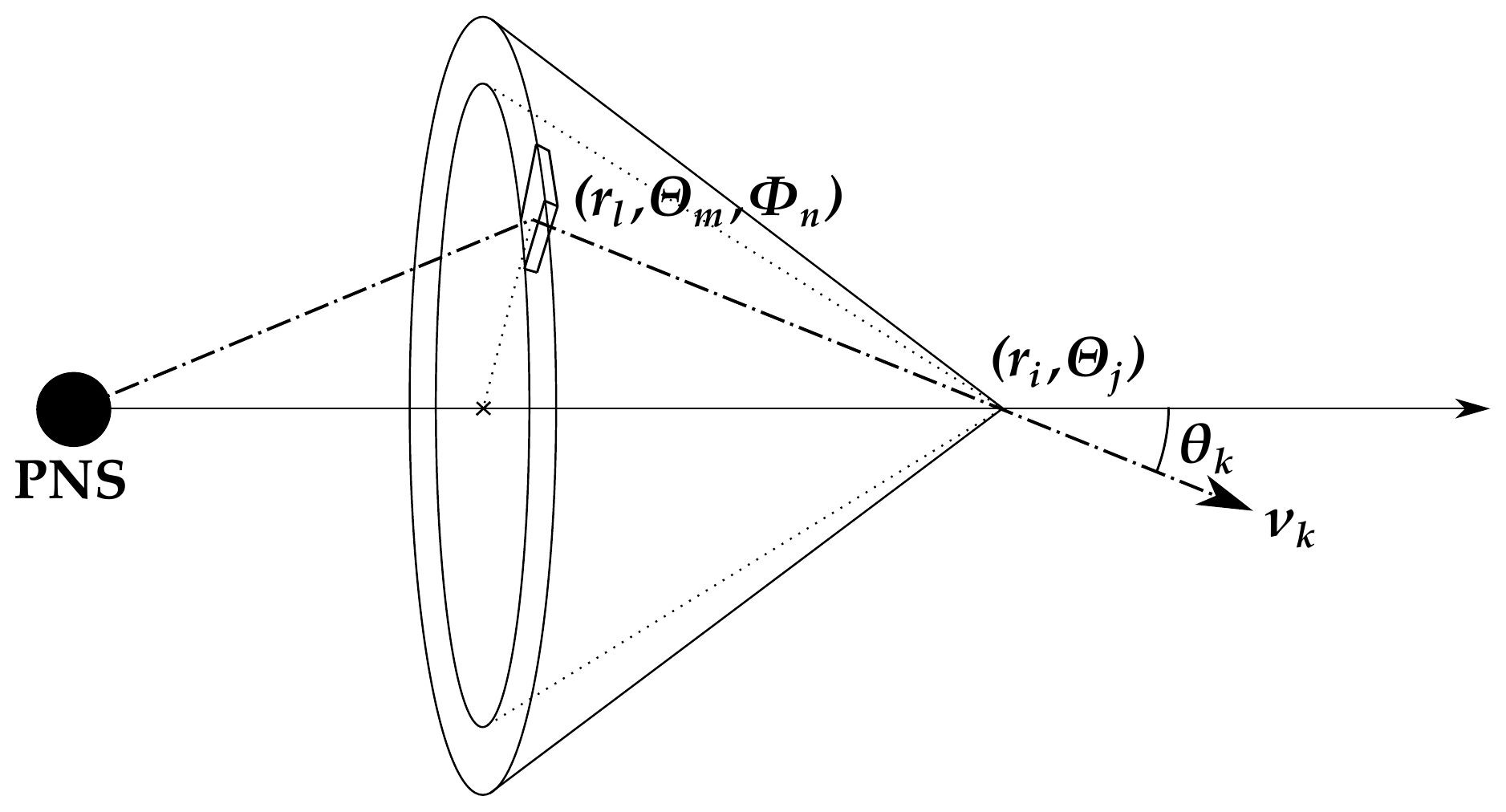}
	\caption{Schematic description for equation (\ref{halomaster}).
	Emitted neutrinos are scattered at a position $(r_l,\Theta_m,\Phi_n)$ towards position $(r_i,\Theta_j)$ with the local radial intersection angle $\theta_{[i,j],[l,m,n]}$.
	Neutrinos with the local polar angle $\theta_k$ at $(r_i,\Theta_j)$ are given by summation over $(l,m,n)$ along trajectory $\theta_k$.
	}
	\label{fig:halo_eq7}
\end{figure}

So far we have calculated the source distribution functions of the halo neutrino population for the original 2D CCSNe simulation.  Now it is necessary to convert that information into a 3D volume of sources to calculate how scattering from each zone contributes halo neutrinos at wide angles to the overall neutrino number densities contained in the original 2D simulation.  Shown in Figure~\ref{fig:halo_eq7} is a diagram of the geometry of neutrino sources which must be summed over.  For each radial and polar angle zone in the 2D simulation, indexed $i$ and $j$ respectively, we create a 3D axisymmetric clone of the 2D data set around the polar axis with $N_{\Phi} \sim 2 N_{\Theta}$.  We will define the indices $l,\ m,\ n$ to be the radial, polar angle, and azimuthal angle bins, respectively, of the cloned 3D set of halo neutrino sources.  Presented as a discrete sum, the energy and angular number density distribution of the halo neutrinos in each radial and polar angle zone, $i,j$, is then,
\begin{equation}
\rho^{\rm halo} \left( \alpha, E_\nu,\theta_k,r_i,\Theta_j \right)  = \sum_{l,m,n} \Omega_{[i,j],[ l,m,n ]}  \Pi\left( \theta_k; \theta_{[i,j],[l,m,n]}, \delta_{[i,j],[l,m,n]}/2\right)\Delta \theta_k \rho^{\rm halo} \left( \alpha, E_\nu,r_l,\Theta_m \right) \, ,
\label{halomaster}
\end{equation}
with $\theta_k$ as the local polar angle relative to the unit normal in zone $\{r_i,\Theta_j\}$,
\begin{equation}
\theta_{[i,j],[l,m,n]} = \cos^{-1} \left( \hat r_{i,j}\cdot \left( \vec r_{i,j} - \vec r_{l,m,n} \right) / \vert  \vec r_{i,j} - \vec r_{l,m,n} \vert \right) \, ,   
\end{equation} 
representing the local radial intersection angle in zone $[i, j]$ of the emission from zone $[l,m,n]$.  The emitting zone has finite angular size, $\delta_{[i,j],[l,m,n]}$, as viewed from the target zone, so we must split the incident neutrino number density across a range of $\theta_k$ bins, centered on $\theta_{[i,j],[l,m,n]}$.  We approximate this distribution to be uniform in the local azimuthal coordinates, so the total number density of neutrinos arriving in the $\theta_k$ angular bin is reduced by a factor of $ \Pi\left( \theta_k; \theta_{[i,j],[l,m,n]}, \delta_{[i,j],[l,m,n]}/2\right)\Delta \theta_k$, where $\Pi(x;a,b)$ is the boxcar function on the interval $a \pm b$, normalized so that the sum $\sum_k \Pi\left( \theta_k; \theta_{[i,j],[l,m,n]}, \delta_{[i,j],[l,m,n]}/2\right)\Delta \theta_k \equiv 1$.   The relative flux dilution of halo sources is accounted for by the solid angle term, $\Omega_{[i,j],[ l,m,n ]}$, which is the solid angle subtended by the target zone, $[i,j]$, relative to the source zone $[l,m,n]$, assuming that all zones are roughly spherical with radius $\mathcal R_{l,m,n} = (3 V_{l,m,n}/4\pi)^{1/3}$ and $\mathcal R_{i,j} = \mathcal R_{l=i,m=j,n=0}$.

Equation (\ref{halomaster}) makes the $4$ dimensional nature of the halo neutrino population apparent. Within all zones $\{r_i,\Theta_j \}$, there is a distribution of neutrino number density with respect to neutrino energies, $E_\nu$, and angle relative to the local radial direction, $\theta_k$. By evaluating Equation (\ref{halomaster}) everywhere within the envelope and recombining it with the radially emitted neutrino densities along a given ray, we have an initial map of the total neutrino distribution within the envelope.

We now need to reduce this information into quantities which are germane to solution of the collective neutrino oscillation EOM.  To begin with, we will suppress the notation such that the total neutrino number density distribution in the envelope is
\begin{equation}
\rho_{\nu_\alpha} (\theta,r_i,\Theta_j ) = \sum_{E_\nu} [\rho^{\rm halo}\left( \alpha, E_\nu,\theta,r_i,\Theta_j \right) + \rho^{\rm ray} \left( \alpha,E_\nu,\theta,r_i,\Theta_j \right)] \label{total},
\end{equation}
summing over neutrino energies and leaving the dependence of $\rho_{\nu_\alpha} (\theta,r_i,\Theta_j  )$ on $E_\nu$ and $\alpha$ implicit.  From this starting point we can calculate the $\nu-\nu$ forward scattering contributions to the collective neutrino oscillation equations of motion directly.  Note that we treat the $\theta$ distribution for $\rho^{\rm ray}$ by requiring that it is distributed with uniform intensity on the surface of the neutrino sphere, as the neutrinos in $\rho^{\rm ray}$ have not undergone any direction changing scattering, but this does not create any constraint that $\rho^{\rm halo}$ be defined in terms of any pseudo-emission surface.  Solving Equation~(\ref{total}) thus allows the specification of $\rho_{\nu_\alpha} (\theta,r_i,\Theta_j )$ for all azimuthal angles $0<\theta<\pi$ in each of the simulation zones $\{r_i,\Theta_j \}$. For reasons that will become clear shortly, we will split these Hamiltonian contributions into two pieces,
\begin{eqnarray}
H^{\rm out}_{\nu\nu}\left(r_i,\Theta_j \right) = \Sigma_{\alpha} \sqrt{2}G_{\rm F}\int^{0}_{1} \left( 1-\cos\theta_{\rm ref}\cos\theta \right) \left[ \rho_{\nu_\alpha} (\theta,r_i,\Theta_j ) - \bar\rho_{\nu_\alpha} (\theta,r_i,\Theta_j ) \right] d\cos\theta \, ,\notag\\
\label{Hout}
\end{eqnarray}
and
\begin{equation}
H^{\rm in}_{\nu\nu}\left(r_i,\Theta_j \right) = \Sigma_{\alpha} \sqrt{2}G_{\rm F}\int^{-1}_{0} \left( 1-\cos\theta_{\rm ref}\cos\theta \right) \left[ \rho_{\nu_\alpha} (\theta,r_i,\Theta_j) - \bar\rho_{\nu \alpha} (\theta,r_i,\Theta_j) \right] d\cos\theta \, ,
\label{Hin}
\end{equation}
which are the $\nu-\nu$ Hamiltonian contributions for a radially directed neutrino from outward directed neutrinos and inward directed neutrinos, respectively. {Note that the choice of a radially directed reference trajectory is relevant for Equations (\ref{Hout}) and (\ref{Hin}) in that it allows the substitution for the angle of intersecting neutrino beams, $\theta_{\rm int}\rightarrow \theta$, because $\cos\theta_{\rm int} = \cos\theta_{\rm ref}\cos\theta = \cos\theta$ for the radially directed case where $\cos\theta_{\rm ref} = 1$.  This choice is not arbitrary on our part but is two fold.  First, $\cos\theta_{\rm ref} = 1$ is selected to produce the most conservative estimate of $H_{\rm out}$, which is minimized for angles close to $\theta = 0$ where the bulk of the neutrino number density lies due to emission from the core.  Second, choosing $\cos\theta_{\rm ref} = 1$ removes the local polar angle dependence, $\propto \sin\theta_{\rm ref}\sin\theta$, for intersecting neutrino trajectories in the zone $\{r_i,\Theta_j \}$ where we wish to know $H^{\rm in}_{\nu\nu}$ and $H^{\rm out}_{\nu\nu}$.}

Equations~(\ref{Hout}) and~(\ref{Hin}) can be evaluated explicitly under the assumption that the neutrino density matrices $\rho_\nu$ and $\bar\rho_\nu$ are flavor diagonal, i.e., in the absence of neutrino flavor transformation, for all points within the CCSN envelope.  With these constructions we can quantify the amount of $\nu-\nu$ flavor transformation Hamiltonian \lq\lq weight\rq\rq which is flowing inward vs.~outward.  This is a critical step in implementing multi-angle collective neutrino oscillation with the inclusion of the halo population~\cite{Cherry:2013a}, as inward directed neutrinos cannot be accommodated by the numerical methods presently available (although considerable progress has been made toward rectifying this issue for single-angle collective neutrino oscillation calculations~\cite{Cirigliano:2018a,Richers:2019a}).

\begin{figure*}[htbp]
	\begin{minipage}{0.5\hsize}
	    \centering
        \includegraphics[width=0.8\linewidth]{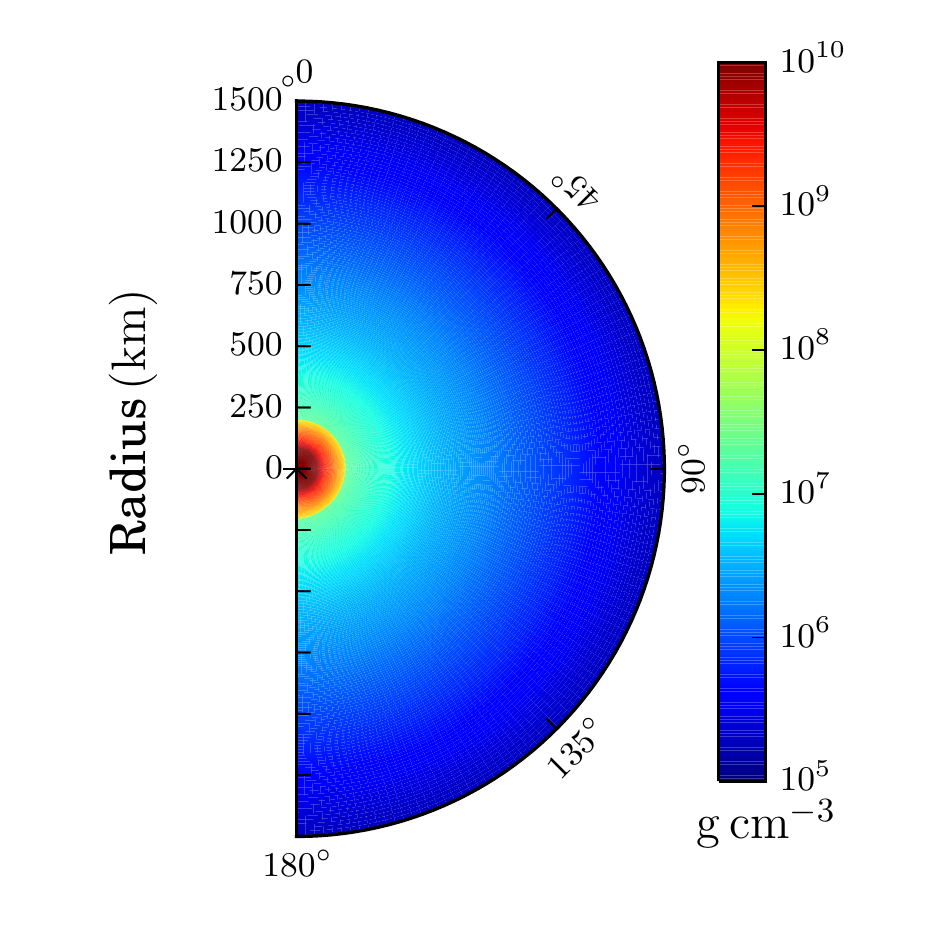}
	\end{minipage}
	\begin{minipage}{0.5\hsize}
	    \centering
        \includegraphics[width=0.8\linewidth]{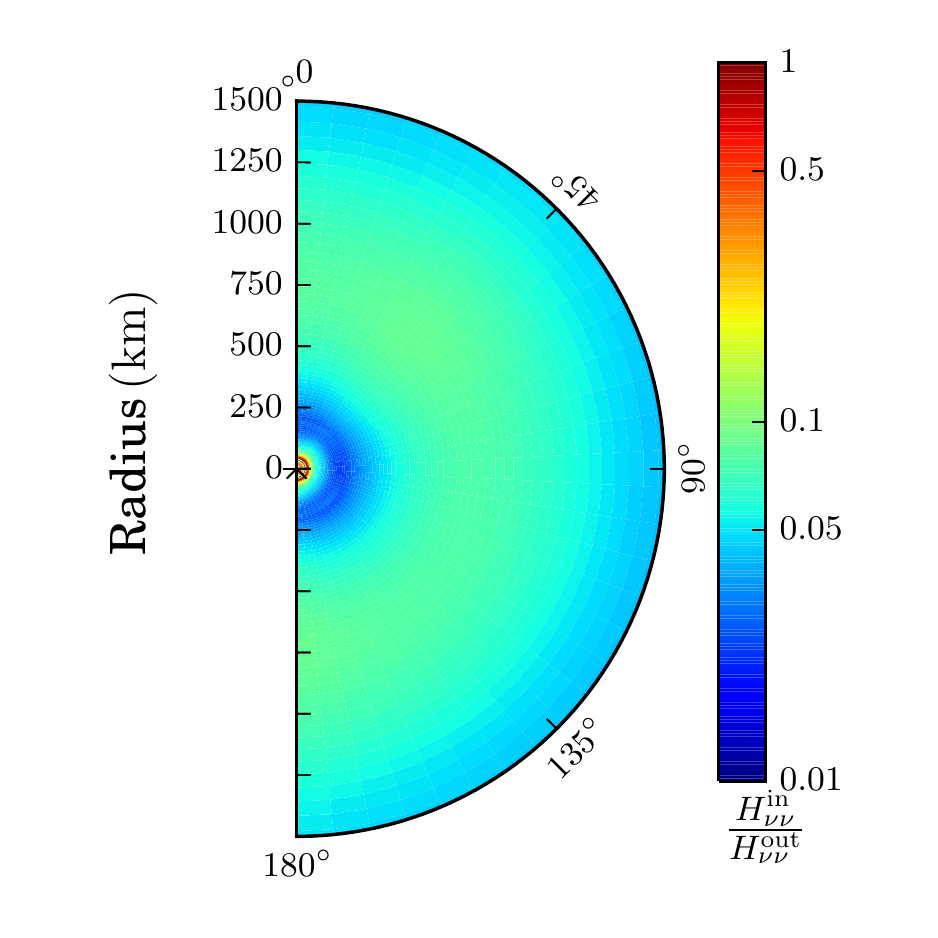}
	\end{minipage}

	\begin{minipage}{0.5\hsize}
	    \centering
        \includegraphics[width=0.8\linewidth]{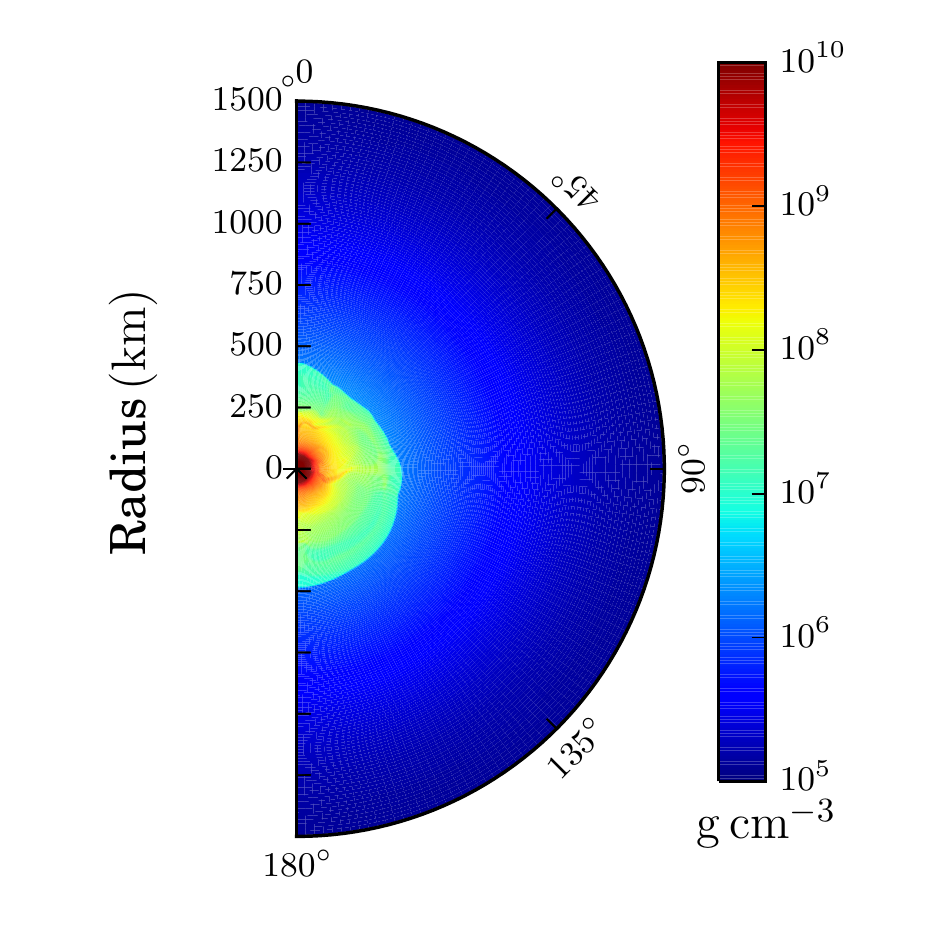}
	\end{minipage}
	\begin{minipage}{0.5\hsize}
	    \centering
        \includegraphics[width=0.8\linewidth]{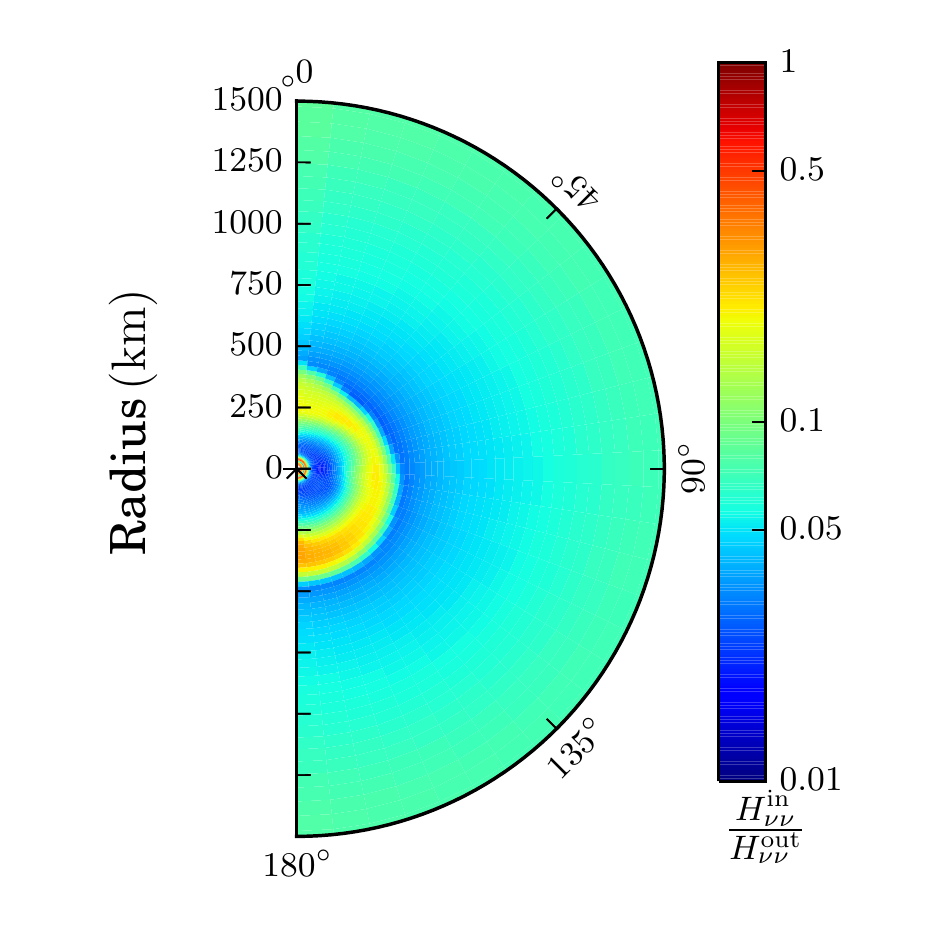}
	\end{minipage}

	\begin{minipage}{0.5\hsize}
	    \centering
        \includegraphics[width=0.8\linewidth]{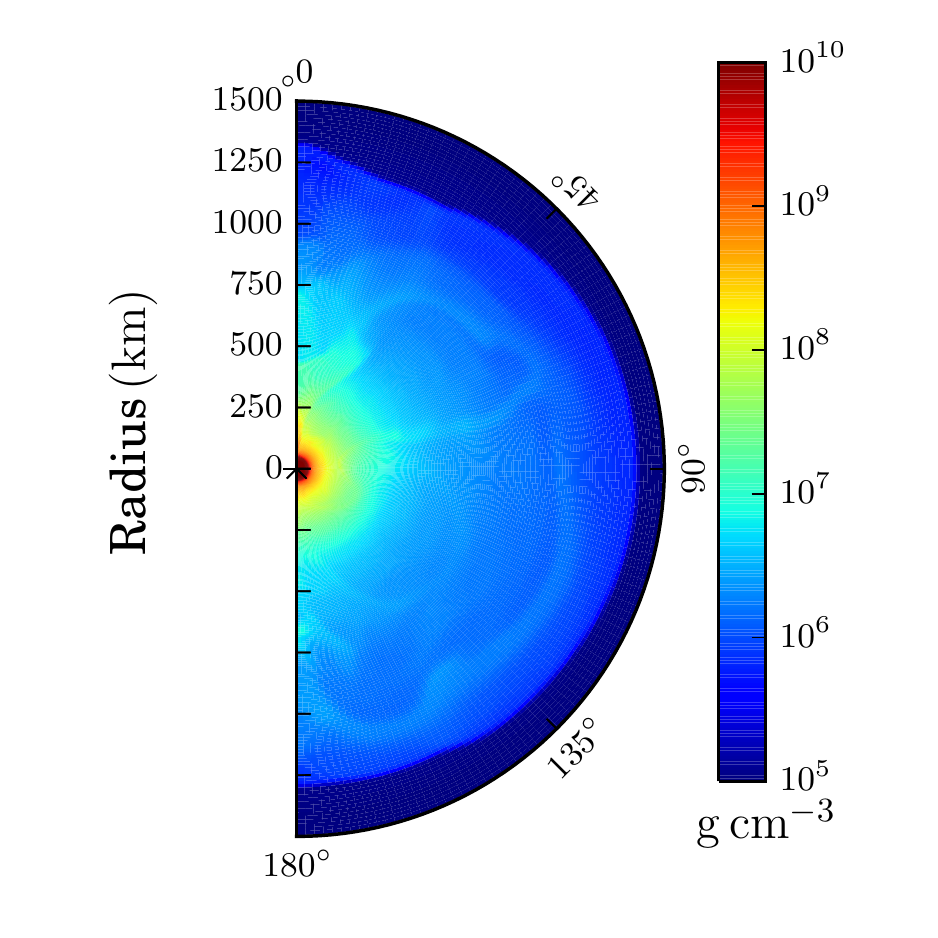}
	\end{minipage}
	\begin{minipage}{0.5\hsize}
	    \centering
        \includegraphics[width=0.8\linewidth]{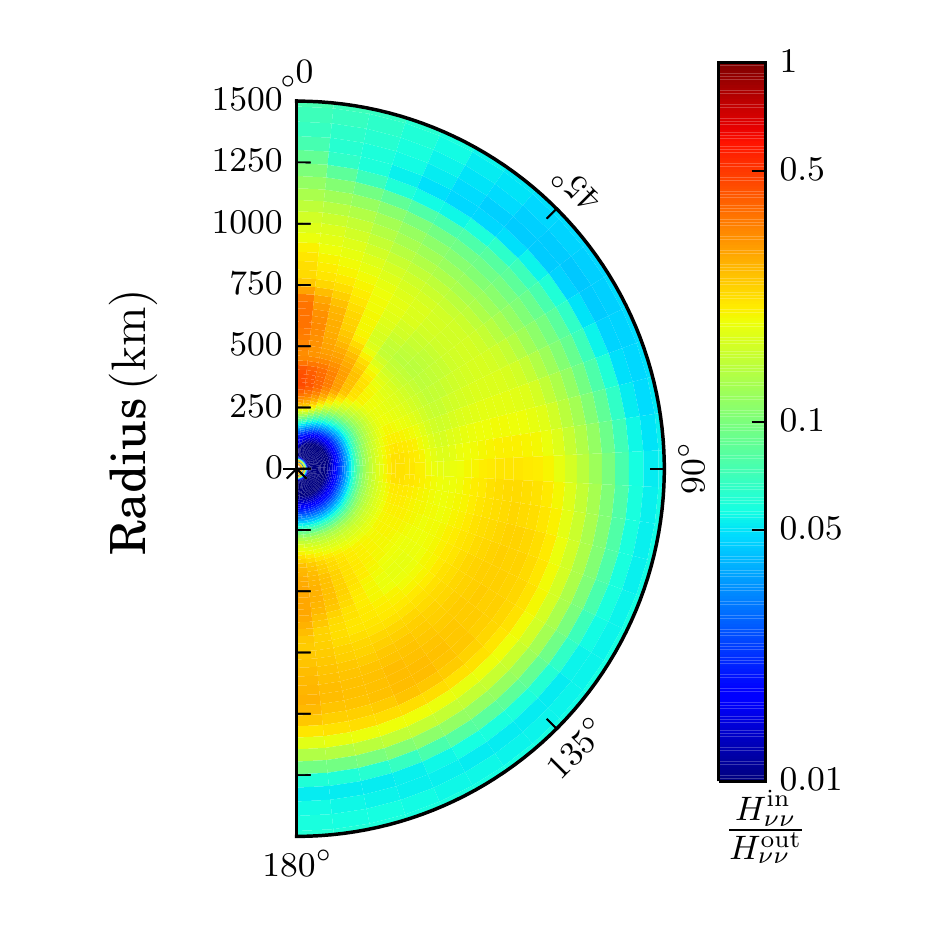}
	\end{minipage}
	\caption{Density profile (left) and Halo contribution (right).
	These figures show three time snapshots, postbounce time $86\, \rm ms,\ 136\, \rm ms$, and $186\, \rm ms$, from top to bottom.
	Color scale for the right panels indicates the ratio of the self-interaction Hamiltonian of inward contribution to outward one.
	An region where this ratio exceeds $0.1$ should not be treated by the bulb+halo model.
	A movie presenting the time evolution is available at http://tron.astron.s.u-tokyo.ac.jp/\texttt{\char`\~}zaizen/halo-movie.html
	}
	\label{fig:Density-Halo}
\end{figure*}

To create a quantitative metric on the suitability of our multi-angle collective neutrino oscillation calculations, we require that the ratio $H^{\rm in}/H^{\rm out}$ be less than~$10\%$ at all radii along which we solve Equation~(\ref{diffevol}).  This guarantees that the portion of the halo population which cannot be included in the multi-angle collective neutrino oscillation calculations is, at most, a sub-leading order contribution to $H_{\rm E,u}$.  Mapping the ratio $H^{\rm in}/H^{\rm out}$ for all points within the envelope is the first step in our safety checks.  Figure~\ref{fig:Density-Halo} shows the results of this procedure for selected time snapshots.

The second step is to check that the flavor diagonal condition used in evaluating Equations~(\ref{Hout}) and~(\ref{Hin}) is valid up to the radius at which we plan to begin the full collective neutrino oscillation calculation.  Beneficially, the large matter densities in the inner regions of the CCSN envelope suppress neutrino flavor conversion at small radii.  To establish the maximum radius at which we can perform our full collective neutrino oscillation calculation, we perform a collective neutrino oscillation calculation which omits the halo neutrinos.  The results of this calculation are sufficient to establish the initial radius, $r_{\rm init}$, below which the flavor diagonal assumption is valid due to matter density suppression of collective neutrino oscillation.

The third preparatory step is to verify that both of the above conditions are satisfied simultaneously along the trajectory where we calculate collective neutrino oscillation.  While the ratio $H^{\rm in}/H^{\rm out}$ may grow larger than~$10\%$ at some radii along a given trajectory, so long as those locations are within the region of matter suppression of collective neutrino oscillation we consider the effects of the halo neutrino population on flavor conversion to be negligible.  Under the condition that $H^{\rm in}/H^{\rm out} < 10\%$ at all radii greater than $r_{\rm init}$ we consider a trajectory safe for the full calculation of collective neutrino oscillation including all radially emitted neutrinos and outward directed halo neutrinos.

Once we are satisfied that the trajectory we are considering is suitable for collective neutrino oscillation calculations the fourth and final step is to return to the neutrino distribution map we have created by summing $\rho^{\rm halo}$ and $\rho^{\rm ray}$ to set the initial conditions for our collective neutrino oscillation calculation
({e.g., Equation (\ref{total}}) with $0\leq \theta_{k}\leq \pi/2$).
Similarly to Ref.~\cite{Cherry:2013a}, we generate our initial condition on a pseudo-emission surface which we call the halosphere, with radius, $R_H$, taken to be $15 \%- 20\%$ less than the measured onset of collective neutrino oscillation at $r_{\rm init}$.  Note that this step does force us to truncate the halo neutrino number density as a function of angle ($\theta < \pi/2 $) and limit the range of solutions to outward directed trajectories.  Energy and emission angle distribution of neutrinos on the surface of the halosphere is taken directly from the local distribution of neutrinos in the envelope at that point $\rho_{\nu_\alpha} \left(E_\nu,\vartheta\right)\vert_{r=R_H} = \rho^{\rm ray}\left( \alpha,E_\nu,\theta,r_i = R_H,\Theta_j \right) + \rho^{\rm halo}\left( \alpha,E_\nu,\theta,r_i = R_H,\Theta_j \right)$.  While this description of the initial condition is straightforward, numerical convergence of collective neutrino oscillation calculations are difficult to achieve unless the angular bins of $\rho_{\nu_\alpha} \left(E_\nu,\vartheta\right)\vert_{r=R_H}$ are chosen very carefully.  We discuss our approach and technical details of that selection in the Appendices.

\subsection{Collective neutrino oscillation}
Now we move on to describe how to implement the fourth procedure in the previous section.
The flavor conversions including collective neutrino oscillation are performed by using the bulb model as a starting point \cite{Duan:2006a, Fogli:2007a, Dasgupta:2008a}, and extending the formalism to include halo neutrinos \cite{Cherry:2013a}.
The traditional bulb model requires uniform and isotropic neutrino emissions under environments depending only on radius $r$.
On the other hand, the bulb+halo model takes initial conditions with anisotropic angular distribution due to the neutrino-nucleus scattering.
Our calculations consider initial neutrino flux with the wider intersection angle given in section \ref{sec:2_halo}.

Neutrino states are simply given by a density matrix $\rho_{\nu}(r; E_{\nu},\theta)$ with neutrino energy $E_{\nu}$ and angular mode $\theta$ at a radius $r$.
The density matrix $\rho_{\nu}$ includes neutrino distribution in diagonal term
\begin{eqnarray}
\mathrm{diag}(\rho_{\nu}) = \left(\rho_{\nu_e}, \rho_{\nu_{\mu}}, \rho_{\nu_{\tau}}\right).
\end{eqnarray}
In the bulb+halo model, we adopt a total neutrino distribution as the initial condition,
 \begin{eqnarray}
\rho_{\nu_{\alpha}} = \rho^{\mathrm{ray}}_{\nu_{\alpha}} + \rho^{\mathrm{halo}}_{\nu_{\alpha}},
\end{eqnarray}
 {which we can obtain from the first to third procedure in the previous section}.  
This halo flux $\rho^{\mathrm{halo}}$ provides broader angular distribution than the bulb emission, and the inclusion of this term is different between the no-halo case and the with-halo case.
In the bulb+halo model, this neutrino distribution is reconstructed to be emitted from the neutrino-halo sphere.
Therefore, the broader intersection angle can be written in the EOM by using the radius of the neutrino-halo sphere $R_H$, not that of the neutrino sphere $R_{\nu}$.
The EOM for a density matrix $\rho_{\nu}$ in a steady state is
\begin{eqnarray}
i\partial_r\rho_{\nu} &&= \left[H_{E,u},~ \rho_{\nu}\right]  \label{diffevol}
\end{eqnarray}
and
\begin{eqnarray}
H_{E,u} = && \frac{1}{v_{r,u}}\left(U\frac{M^2}{2E_{\nu}}U^{\dagger}+\sqrt{2}G_{\mathrm{F}}n_e L \right) \notag \\
&&+ \sqrt{2}G_{\mathrm{F}}\int\mathrm{d}E_{\nu}^{\prime}\mathrm{d}u^{\prime}\left(\frac{1}{v_{r,u}v_{r,u^{\prime}}}-1\right)\left(\rho_{\nu}^{\prime}(E_{\nu}^{\prime},u^{\prime})-\bar{\rho}_{\nu}^{\prime}(E_{\nu}^{\prime},u^{\prime})\right), 
\end{eqnarray}
where $U$ is the Pontecorvo-Maki-Nakagawa-Sakata matrix \cite{PMNS1962}, $M^2$ is a neutrino mass square matrix, and $L$ is $\mathrm{diag(1,0,0)}$.
The radial velocity $v_{r,u}$ with an angular mode $u$ is defined as
\begin{eqnarray}
v_{r,u} &&= \sqrt{1-u\frac{R^2}{r^2}} \\
u &&= \sin^2\theta_{R},
\end{eqnarray}
where $\theta_{R}$ is an emission angle relative to the radial direction on the surface of an emission source.
We treat the emission source as the neutrino sphere $R=R_{\nu}$ in the no-halo case and as the neutrino-halo sphere $R=R_H$ in the with-halo case.
In this calculation, we assume axial symmetry on the neutrino trajectories, that is, we neglect the multi-azimuthal angle effect \cite{Raffelt:2013a}.
This symmetry breaking appears only in the normal mass ordering and the directional azimuth-angle distribution does not affect the flavor conversions in the inverted mass ordering \cite{Mirizzi:2013a, Chakraborty:2014b, Chakraborty:2016a}.

In this work, we choose the following neutrino parameters as in Ref.~\cite{PDG18}: $\Delta m_{21}^2=7.37\times 10^{-5}\mathrm{~eV^2}, \left|\Delta m_{31}^2\right|=2.54\times 10^{-3}\mathrm{~eV^2}, \sin^2\theta_{12}=0.297, \sin^2\theta_{13}=0.0216$, and CP-violation phase $\delta=0$.
We consider only the inverted mass ordering case $\Delta m_{31}^2<0$ because collective neutrino oscillation is suppressed for the normal mass ordering case in our calculation \cite{Esteban:2007a}.
We introduce the rotated state,
\begin{eqnarray}
&&\nu_x = \cos\theta_{23}\nu_{\mu} -\sin\theta_{23}\nu_{\tau} \\
&&\nu_y = \sin\theta_{23}\nu_{\mu} +\cos\theta_{23}\nu_{\tau},
\end{eqnarray}
not the ordinary state $(\nu_e,\nu_{\mu},\nu_{\tau})$ in the following calculations.
This rotation does not affect the electron-type neutrinos, and enables us to understand the three-flavor framework as the combination of two-flavor problems, $\nu_e\leftrightarrow\nu_y$ and $\nu_e\leftrightarrow\nu_x$.
The three-flavor effects associated with $\Delta m_{21}^2$ and $\theta_{12}$ arise from flavor symmetry condition on the total neutrinos fluxes for each flavor, $\Phi_{\alpha}=L_{\nu_{\alpha}}/\langle E_{\nu_{\alpha}}\rangle$, such that $\Phi_{\nu_e} \simeq \Phi_{\bar{\nu}_e} \simeq \Phi_{\nu_x}$, and cause $\nu_e\leftrightarrow\nu_x$ conversion \cite{Friedland:2010a, Dasgupta:2010a, Mirizzi:2011a}.
These effect can emerge even in the normal mass ordering, but neutrino number fluxes in Figure \ref{fig:SN_model} do not satisfy the symmetry condition.
Therefore, $\nu_e\leftrightarrow\nu_y$ conversion is dominant in our calculation.

\begin{figure}[h]
	\centering
    \includegraphics[width=0.6\linewidth]{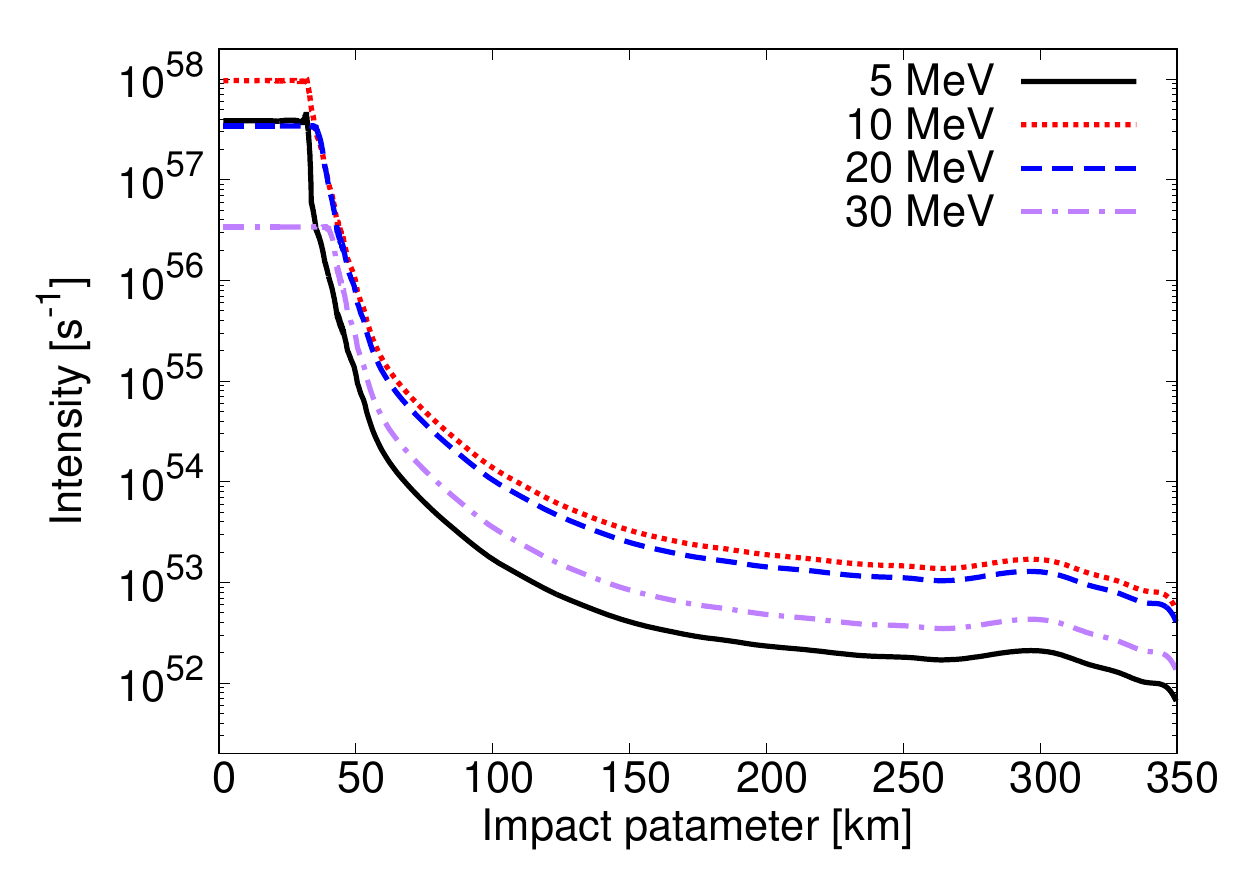}
	\caption{Antielectron neutrino intensity distribution as a function of impact parameter at postbounce time $t_{\rm pb} = 136\mathrm{~ms}$.
	}
	\label{fig:imp_intensity}
\end{figure}

We describe the flavor difference with polarization vectors $\mathbf{P}$ and $\overline{\mathbf{P}}$ in solving the EOM numerically.
This polarization vector reflects independent components of the density matrix and we can transform Equation (\ref{diffevol}) from matrix differential equations including complex values to vector differential equations composed only of real values as
\begin{eqnarray}
\rho_{\nu} &&= \frac{1}{3}I_3 +\frac{1
}{2}\mathbf{P}\cdot\mathbf{\Lambda} \\
\partial_r\mathbf{P}(E_{\nu},u) &&= \mathbf{H}_{E,u} \times \mathbf{P}(E_{\nu},u),
\end{eqnarray}
where $\mathbf{\Lambda}$ is a vector of the Gell-Mann matrices.
This three-flavor formalism is based on Ref.~\cite{Dasgupta:2008a, Zaizen:2018a}.
Then, $P_3$ and $P_8$ correspond to the diagonal terms of the density matrix $\rho_{\nu}$ and initial conditions are written as
\begin{eqnarray}
&&P^i_3(E_{\nu},u) = \rho_{\nu_e}(E_{\nu},u) - \rho_{\nu_x}(E_{\nu},u) \\
&&P^i_8(E_{\nu},u) = \frac{\rho_{\nu_e}(E_{\nu},u) +\rho_{\nu_x}(E_{\nu},u) -2\rho_{\nu_y}(E_{\nu},u)}{\sqrt{3}}.
\end{eqnarray}
If neutrino emission is assumed to be isotropic in the traditional bulb model, these initial conditions depend only on energy distribution.
On the other hand, if we consider neutrino scattering, initial conditions have angular distribution including a halo flux.
Figure \ref{fig:imp_intensity} expresses neutrino intensity distribution over impact parameter $b = R_H\sin\theta_R$ at four select energies at postbounce time $t_{\rm pb} = 136\mathrm{~ms}$.

This energy and angular distribution is calculated from coherent elastic neutrino scattering with background matter, and it is set as the initial conditions in the with-halo case.
The bulb+halo model is set by reconstructing neutrino distribution on the neutrino-halo sphere.
We calculate collective neutrino oscillation with/without halo effects along the northern pole direction of the two-dimensional iron-CCSN model.

\section{Results}
\label{sec:3}

\subsection{Magnitude of Halo}
We show density profile and halo contribution at $86\, \rm ms ,\ 136\,\rm ms$, and $186\, \rm ms $ in Figure ~\ref{fig:Density-Halo}.
This halo contribution expresses the ratio $H^{\rm in}/H^{\rm out}$ of the self-interaction Hamiltonian of inward contribution to outward one.
In regions where this ratio is large, the inward-scattered neutrino flux cannot be neglected.
This dangerous region especially expands over the whole map at $186\mathrm{~ms}$.
However, we find an escape route along $45$ degree direction.
This gap is produced by two-dimensional density structure and neutrino flux distribution. 
Comparing this region with the density profile, we find that the halo contribution is larger inside the shock wave and steeply decreases outside.
This feature is similar to the steep density gradient in a O-Ne-Mg CCSN.
Therefore, the density gradient of the shock wave is having the same effect as the envelope density gradient in O-Ne-Mg CCSNe.
We can thus calculate collective neutrino oscillation safely, ignoring inward-going neutrino flux outside the shock wave.

In this case, we replace the neutrino sphere $R_\nu$ with a neutrino-halo sphere $R_{H}$ as an emission source.
The radius of neutrino-halo sphere almost corresponds to the shock wave location and the flavor conversion does not occur inside this neutrino-halo sphere.
Using this bulb+halo model, we investigate the impacts of outward-going neutrino flux on collective neutrino oscillation signals.

\subsection{Halo effects on flavor conversion}
In our model, the shock wave propagates outward within the occurrence region of collective neutrino oscillation.
This propagation affects the density structure in this region as time passes, and changes the behaviors of the multi-angle matter suppression.
We show the time variation of the density profile along the north polar direction at $86\,\rm ms,\ 136\,\rm ms$, and $186\,\mathrm{ms}$ in Figure \ref{fig:north_density}.
Collective neutrino oscillation at $186\mathrm{~ms}$ is completely suppressed due to matter effects.
Even though this epoch is not the accretion phase of the Z9.6 model CCSNe, the mechanism of complete suppression is the same as described in \cite{Sarikas:2012b}, who demonstrated that collective flavor conversion suppression can take place when accounting for the effects of halo neutrinos.

\begin{figure*}[t!]
	\begin{minipage}{0.5\hsize}
	    \centering
        \includegraphics[width=1.0\linewidth]{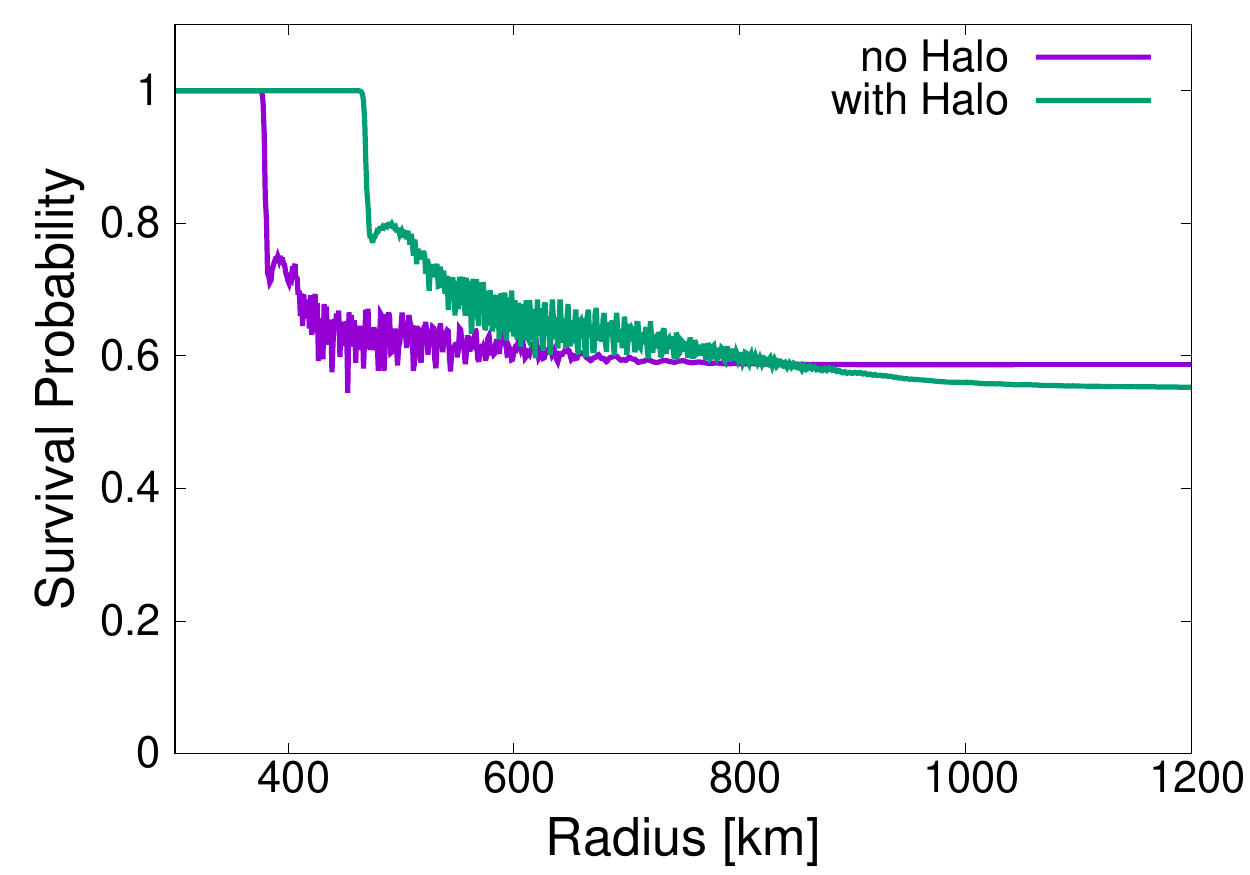}
	\end{minipage}
	\begin{minipage}{0.5\hsize}
	    \centering
        \includegraphics[width=1.0\linewidth]{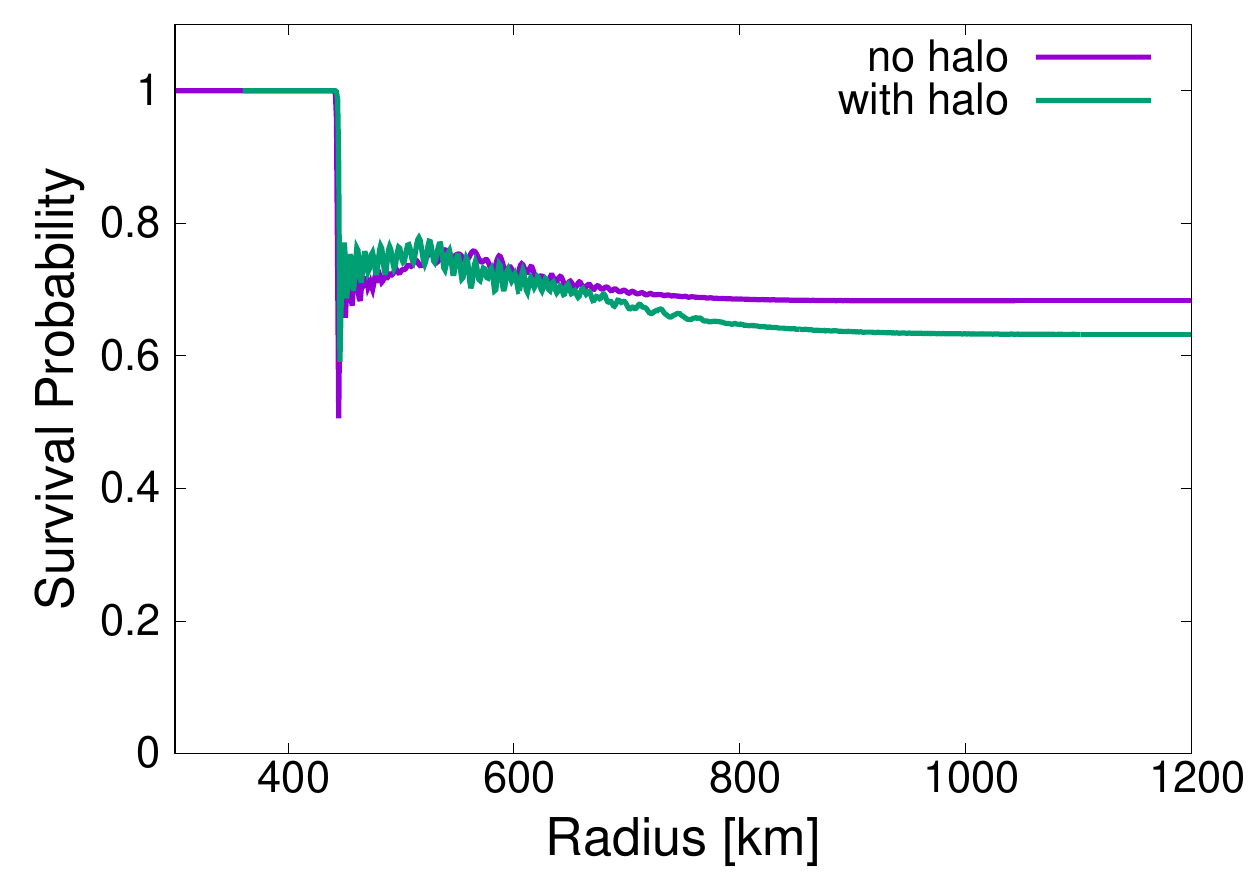}
	\end{minipage}
	\caption{The radial evolution of the energy and angle averaged $\nu_{\rm e}$ survival probability at postbounce time $86\mathrm{~ms}$ (left) and $136\mathrm{~ms}$ (right).
	These survival probabilities are in the inverted mass ordering case.
	}
	\label{fig:surv}
\end{figure*}

We first discuss results at $86\,\rm ms$ and $136\mathrm{~ms}$.
Figure \ref{fig:surv} shows the radial evolution of the survival probability of electron neutrinos averaged over energy and angular mode at $86\,\rm ms$ and $136\,\mathrm{ms}$.
In the no-halo case, at $86\,\mathrm{ms}$, the shock wave is located at $r\sim 200\,\mathrm{km}$ and collective neutrino oscillation occurs outside the shock wave.
The electron density gradually decreases with increasing radius, and flavor conversions occur at $r\sim 400\,\mathrm{km}$.
In the with-halo case, the onset of flavor conversion is clearly delayed.
The with-halo case suggests that the neutrino halo gives additional multi-angle matter suppression for collective neutrino oscillation.
The matter suppression is induced by the phase dispersion due to the different neutrino trajectories.
The maximum value of intersection angle in the no-halo case depends roughly on the inverse square of the radius at a distance $r\gg R_{\nu}$.
On the other hand, halo components have broader angle distribution and give additional phase dispersion.
Halo neutrinos with wide angles strengthen the multi-angle matter suppression, and the onset radius of collective neutrino oscillation is delayed.

At $136\,\mathrm{ms}$, there is no difference between oscillation radius in the no-halo case and the with-halo case.
The shock wave is located at $r\sim 430\,\mathrm{km}$ at this time snapshot.
Neutrinos get free from the matter suppression just after propagating through the shock wave, and collective neutrino oscillation suddenly occurs.
This feature does not change even in the with-halo case.
Therefore, the onset radius of collective neutrino oscillation is the same as in the no-halo case, different from at $86\mathrm{~ms}$.

\begin{figure*}[htbp]
	\begin{minipage}{0.5\hsize}
	    \centering
    	\includegraphics[width=1.0\linewidth]{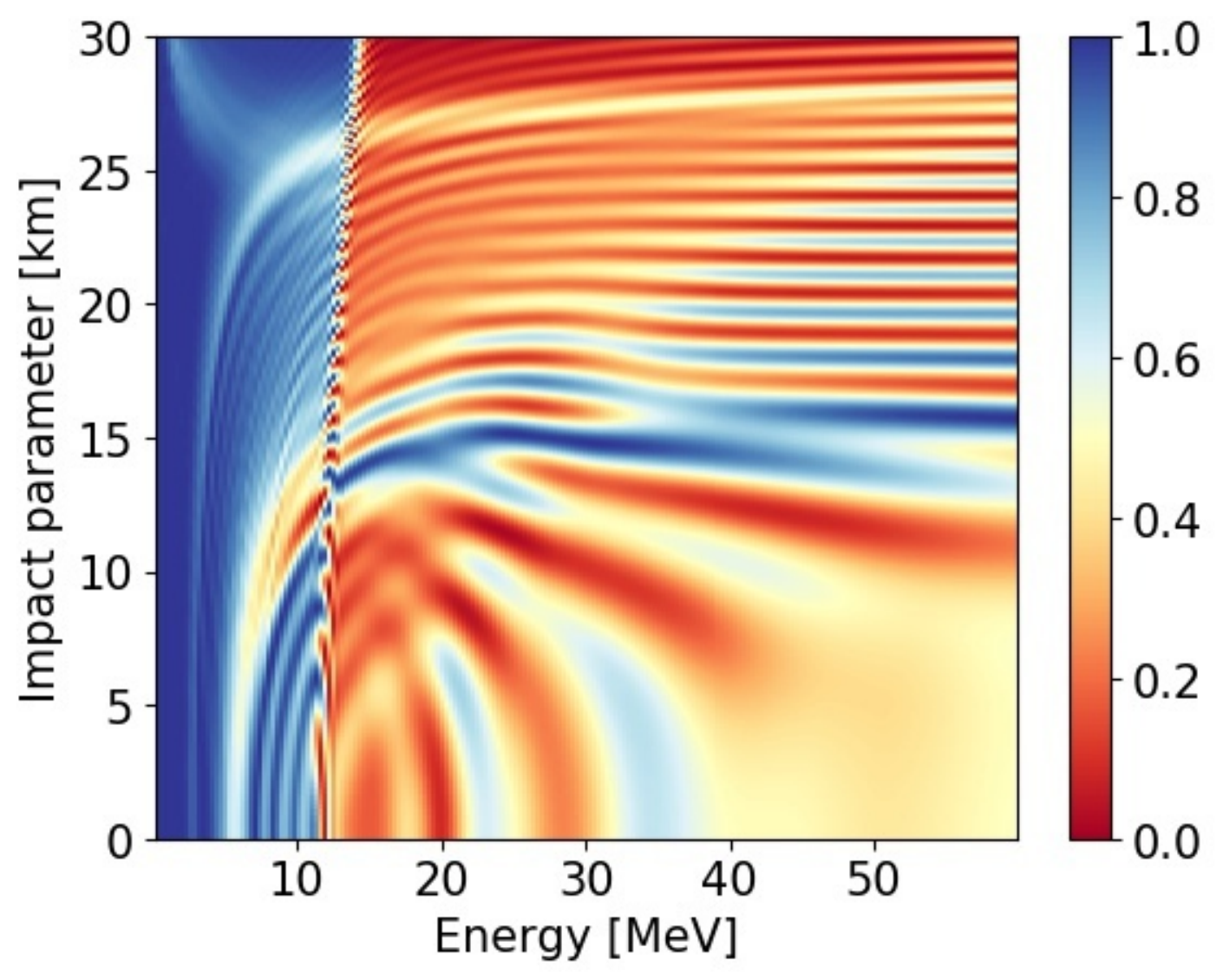}
	\end{minipage}
	\begin{minipage}{0.5\hsize}
	    \centering
    	\includegraphics[width=1.0\linewidth]{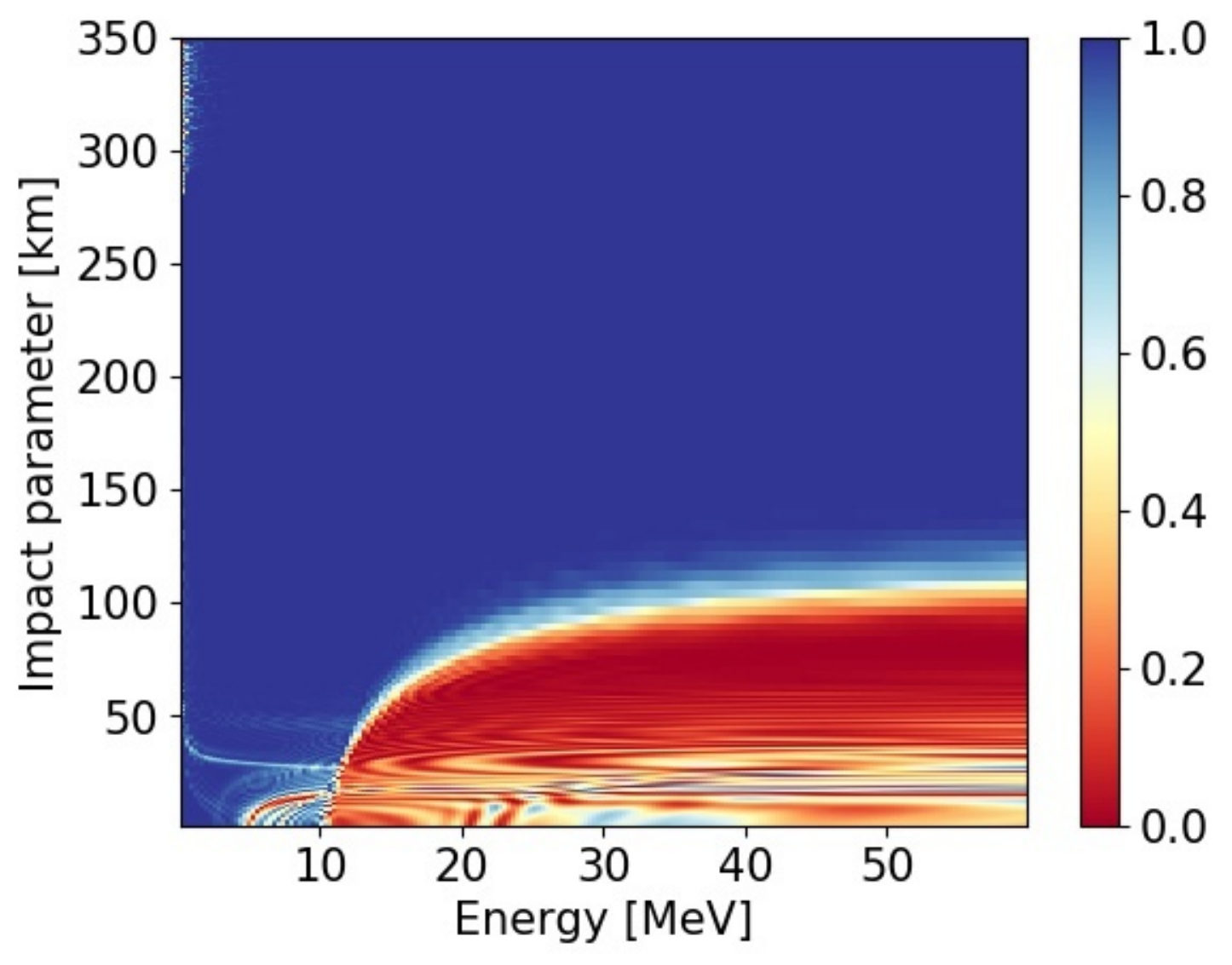}
	\end{minipage}

	\begin{minipage}{0.5\hsize}
	    \centering
    	\includegraphics[width=1.0\linewidth]{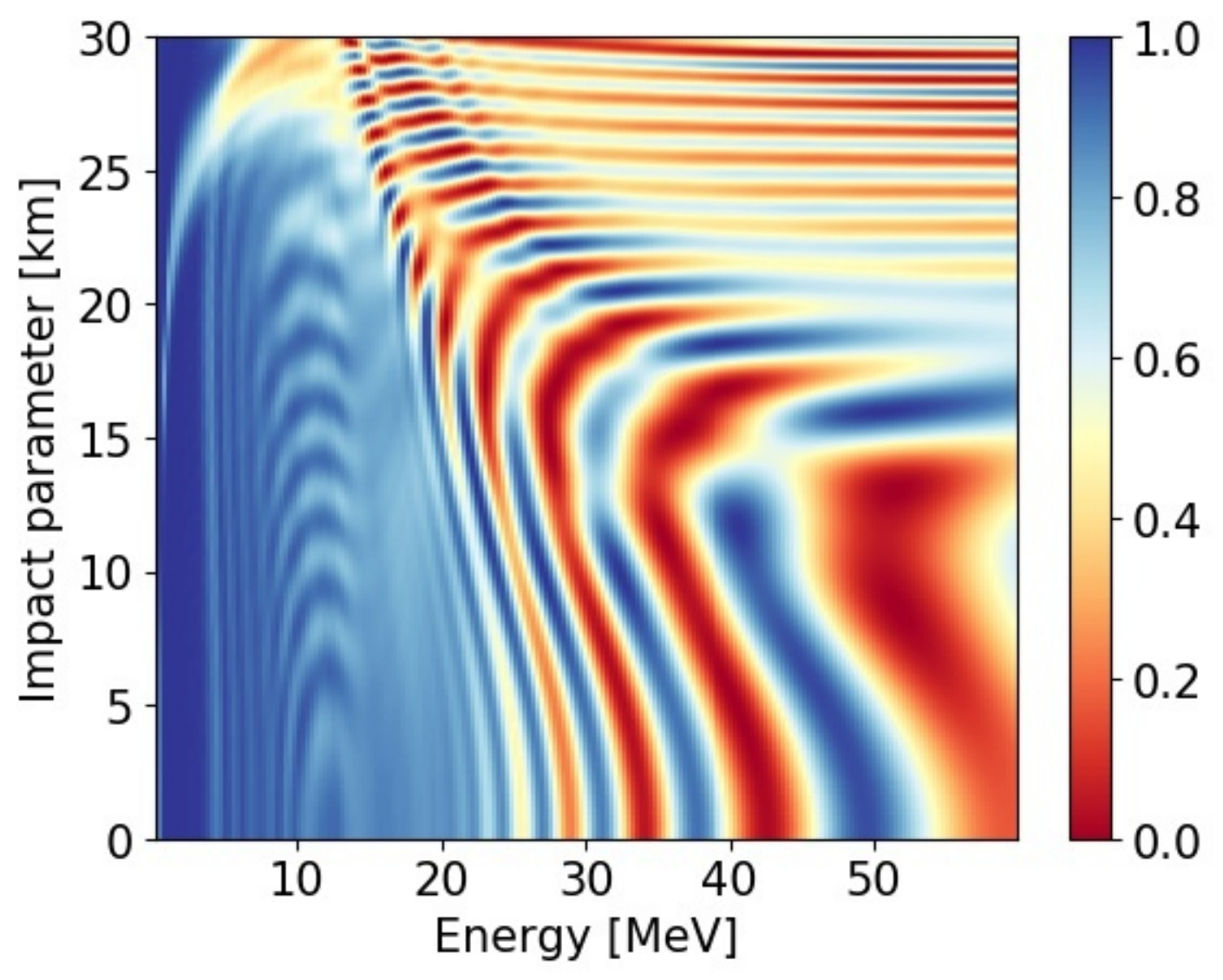}
	\end{minipage}
	\begin{minipage}{0.5\hsize}
	    \centering
    	\includegraphics[width=1.0\linewidth]{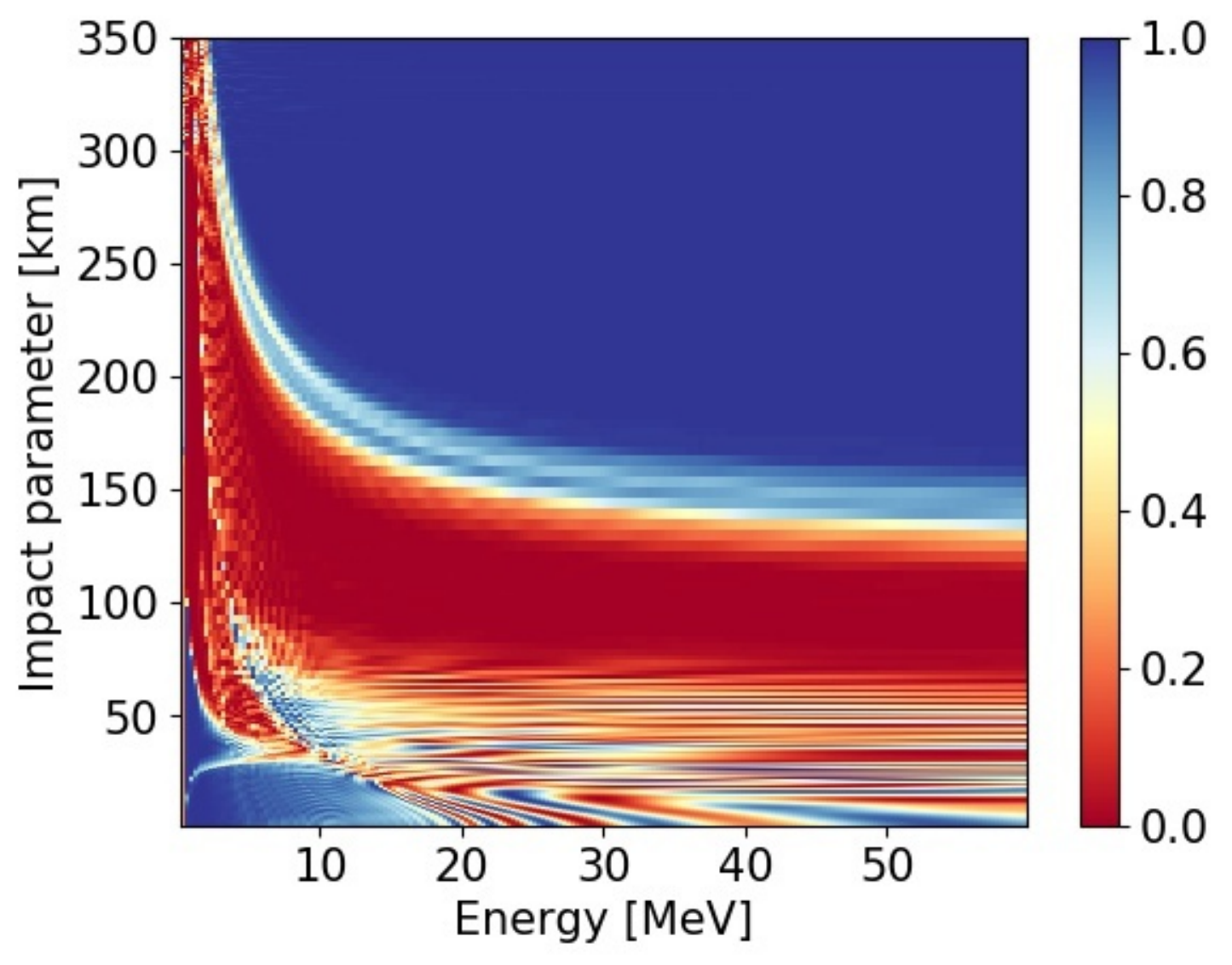}
	\end{minipage}
	\caption{The survival probability contour maps for neutrinos (upper panels) and antineutrinos (bottom panels) on energy-impact parameter plane at $r=1200\mathrm{~km}$ at postbounce time $136\mathrm{~ms}$.
	Left panels are the no-halo case and right panels are the with-halo case.
	}
	\label{fig:uE_contour}
\end{figure*}

Second, we present calculation results of collective neutrino oscillation in the no-halo case and the with-halo case at $136\mathrm{~ms}$ as a representative snapshot.
Figure \ref{fig:uE_contour} shows the survival probability contour maps on energy-impact parameter plane after the collective neutrino oscillation ceases at $r=1200\mathrm{~km}$.

\begin{figure*}[hbtp]
	\begin{minipage}{0.5\hsize}
	    \centering
    	\includegraphics[width=1.0\linewidth]{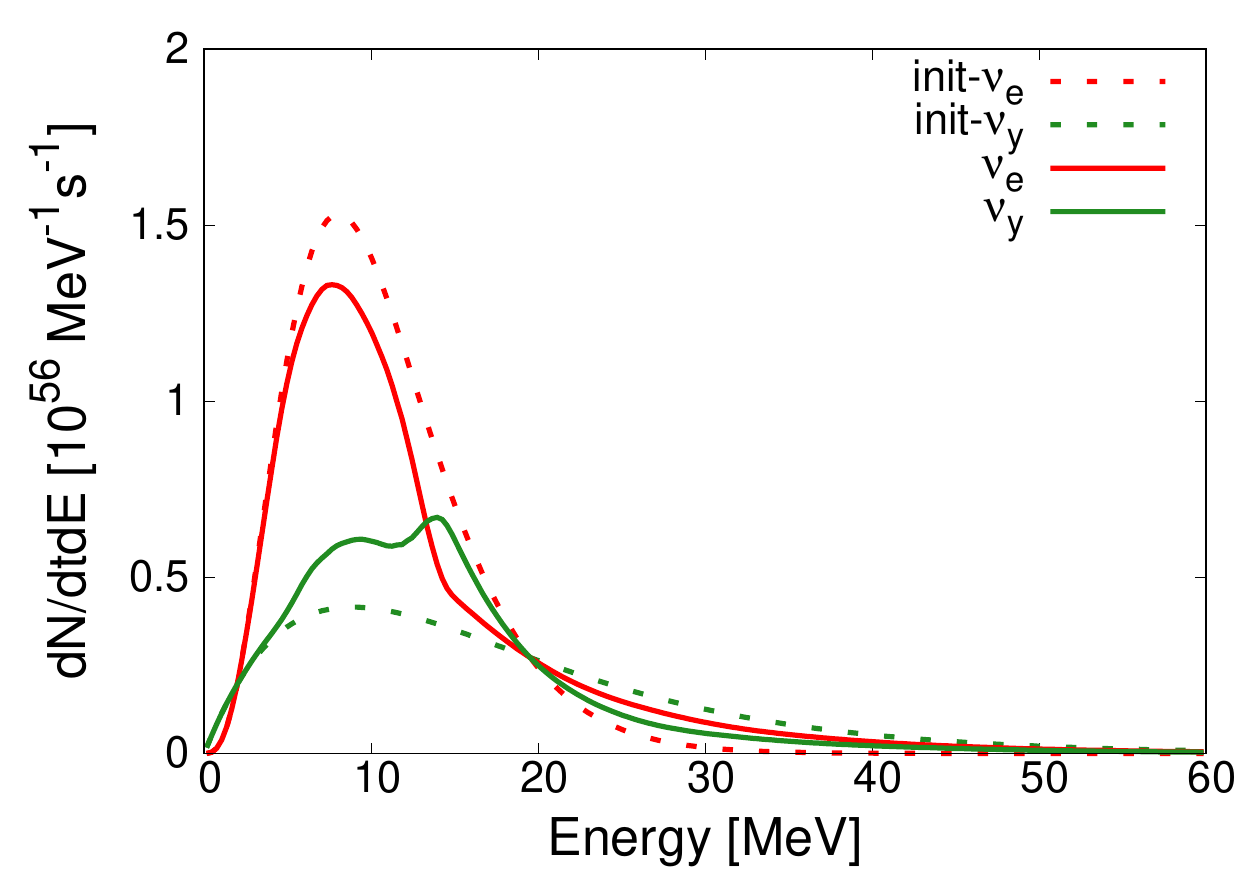}
	\end{minipage}
	\begin{minipage}{0.5\hsize}
	    \centering
    	\includegraphics[width=1.0\linewidth]{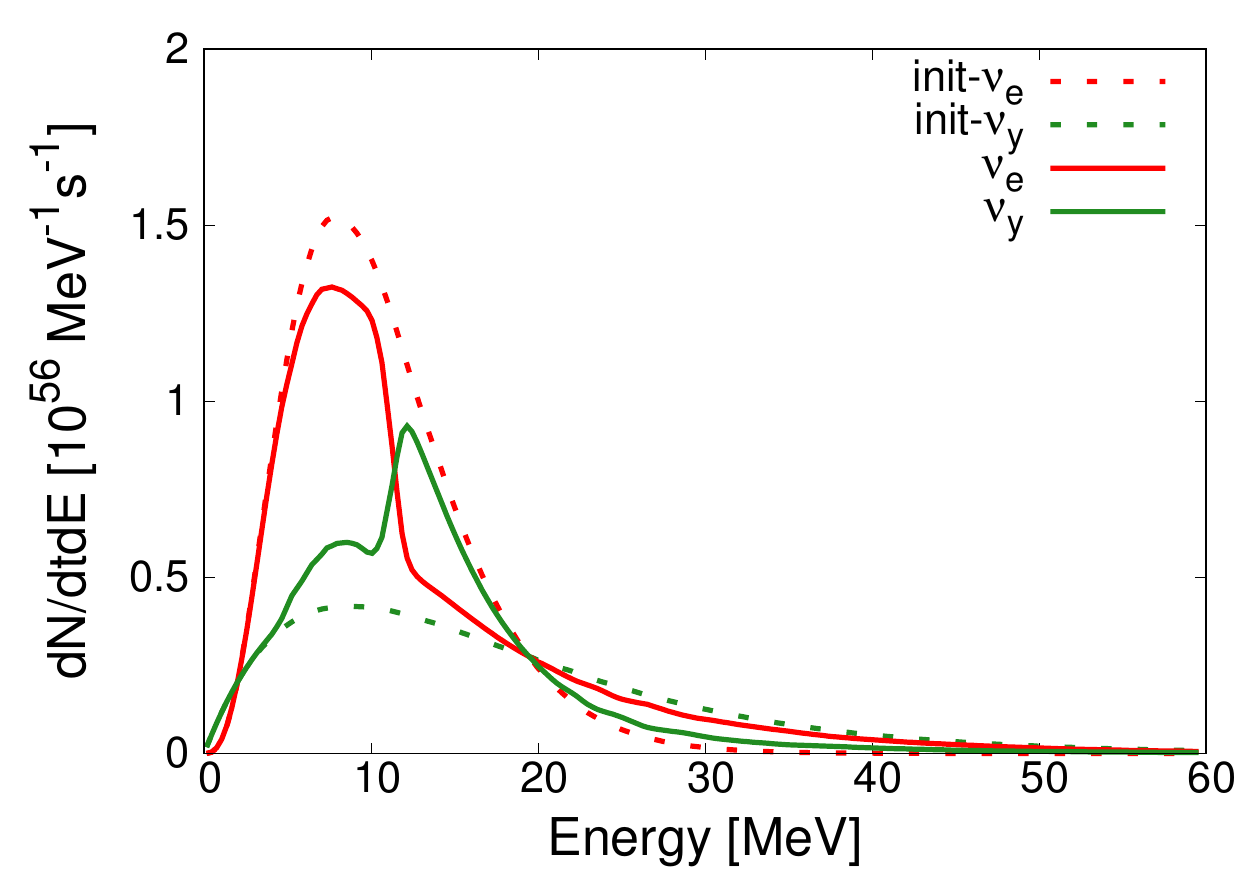}
	\end{minipage}

	\begin{minipage}{0.5\hsize}
	    \centering
    	\includegraphics[width=1.0\linewidth]{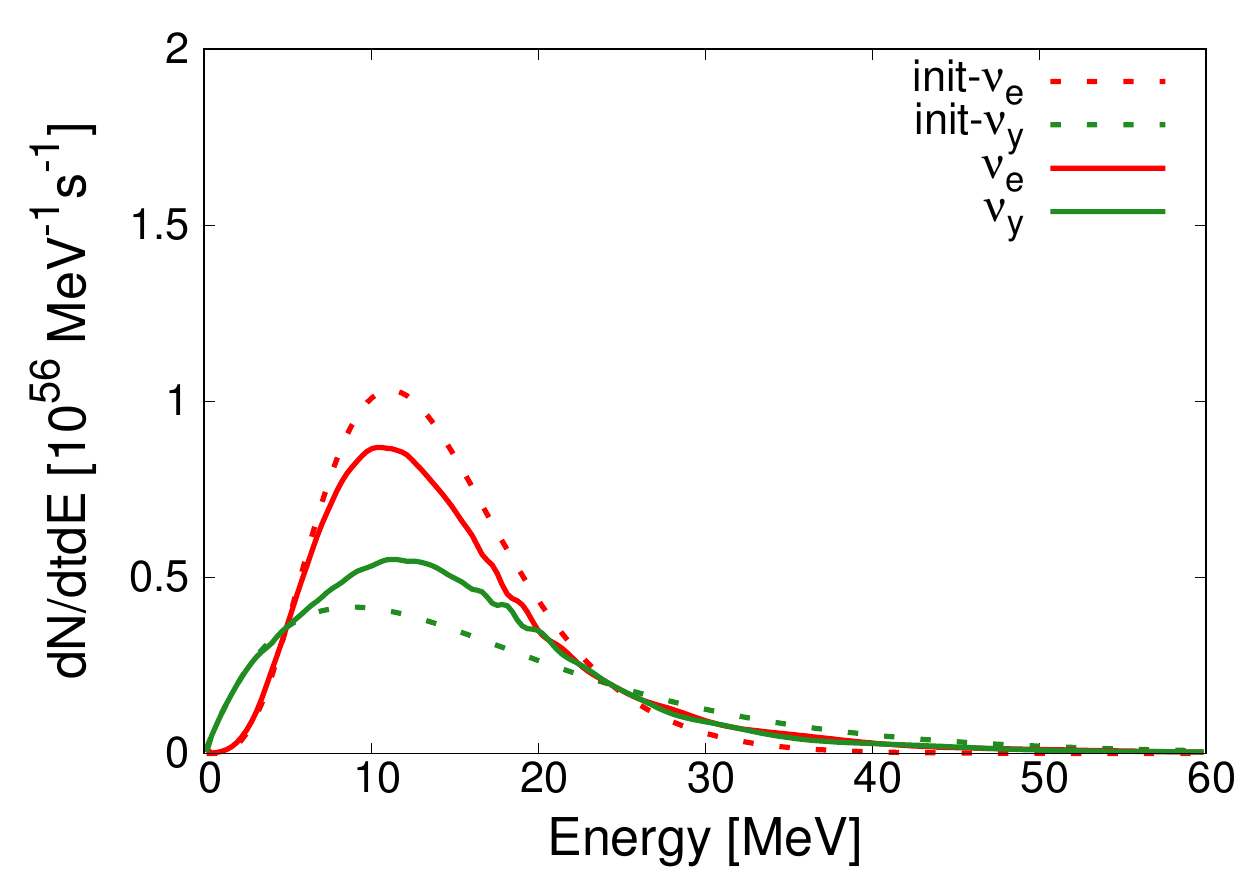}
	\end{minipage}
	\begin{minipage}{0.5\hsize}
	    \centering
    	\includegraphics[width=1.0\linewidth]{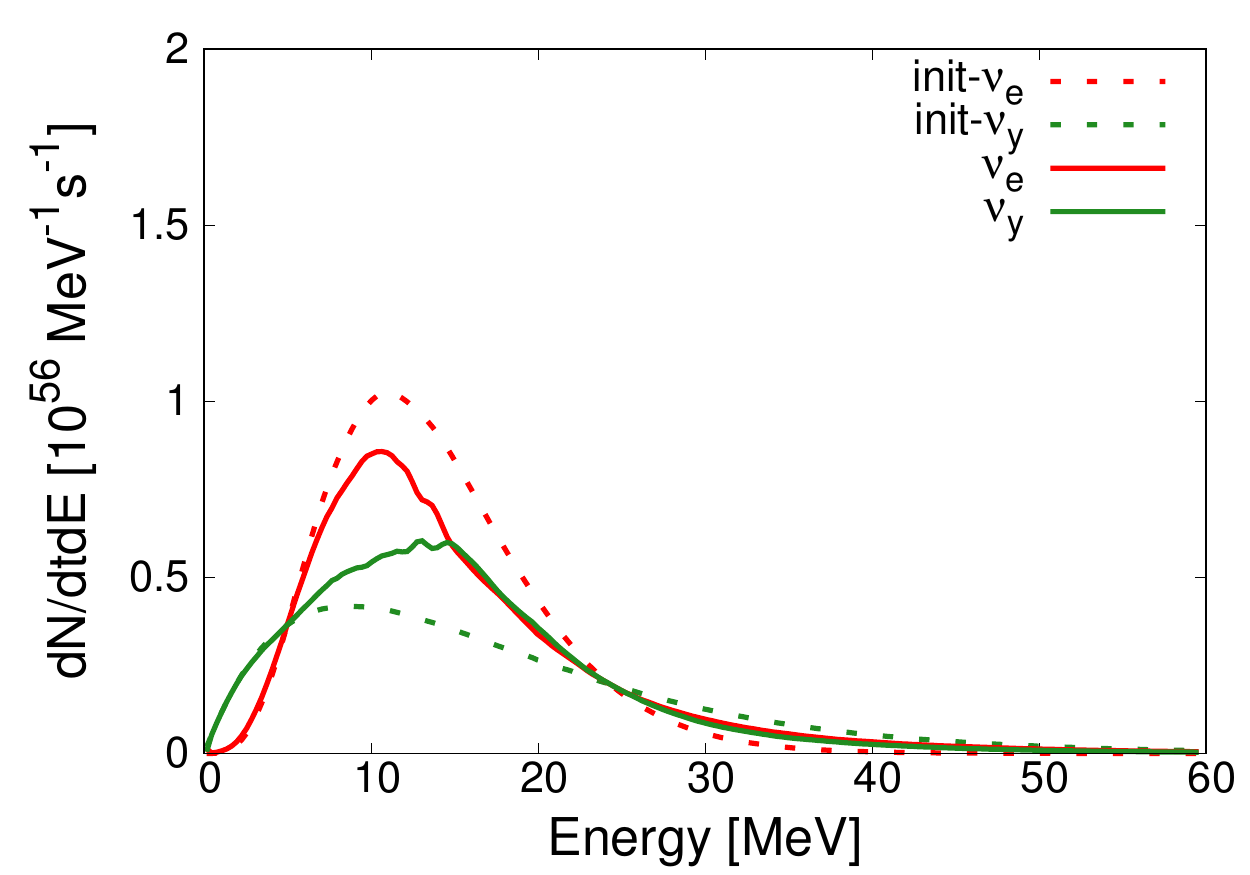}
	\end{minipage}
	\caption{Neutrino spectra (upper panels) and antineutrino spectra (bottom panels) after the collective neutrino oscillation ceases at $r=1200\mathrm{~km}$ at postbounce time $136\mathrm{~ms}$.
	Left panels are the no-halo case and right panels are the with-halo case.
	The $e$ and $y$ flavors are shown in red and green lines, respectively.
	The solid lines are for neutrino spectra after collective neutrino oscillation and the dashed ones for initial spectra.
	}
	\label{fig:spectra}
\end{figure*}

The left panels in
Figure \ref{fig:uE_contour} are results in the no-halo case and the right ones are in the with-halo case.
The top panels show the results for ordinary neutrinos and the bottom ones are for antineutrinos.
The vertical axis expresses the impact parameter $b=r \sin\theta_R$ of emitted neutrinos from neutrino sphere/neutrino-halo sphere which terminates at $r = R_{H}$.  
Emissions with impact parameter beyond the radius of proto-neutron star, $30\mathrm{~km}$, correspond to the contribution from neutrino halo.
The survival probabilities for neutrinos in the no-halo case clearly predict spectral splits as shown in previous works \cite{Duan:2006b}.
In the with-halo case, halo contributions give additional flavor conversions above the low impact parameter region.
These additional oscillations deform the neutrino spectra and affect the detection at Earth.

Figure \ref{fig:spectra} shows the angle-averaged neutrino spectra after collective neutrino oscillation ceases at $r=1200\mathrm{~km}$.  Left panels are the no-halo case and right panels are the with-halo case.
Additional flavor conversion above $\sim 10\mathrm{~MeV}$ occurs, in both neutrinos and antineutrinos.
This feature corresponds to the survival probability contour map in Figure \ref{fig:uE_contour}.
However, it does not appear to match over low energy range below $\sim 10\mathrm{~MeV}$ for antineutrinos.
Halo components in this region show almost complete flavor conversions.
This shift in the flavor transformation pattern is due to the redistribution of halo neutrinos to wider angles.

\subsection{Detectability}
For the Z9.6 model, we find that multi-angle matter suppression prevents collective neutrino oscillation at early and late times, with the exception of a window from $t_{\rm pb} = 70\mathrm{~ms}$ to $t_{\rm pb} = 170\mathrm{~ms}$ during the shock revival epoch of the explosion.  There are two salient questions we would like to answer with regards to the resultant neutrino signal for the Z9.6 simulation: (1) Can we see signatures of the onset / end of collective neutrino oscillation within the received signal?  (2) What impact, if any, does the inclusion of the neutrino halo in the collective neutrino oscillation calculation have on the received signal? 

Inclusion of halo neutrinos tends to sharpen the features of collective neutrino oscillation relative to calculations which omit halo neutrinos.  In the absence of coherently enhanced elastic scattering, the $\nu-\nu$ contribution to Equation (2) is roughly $\propto (R_\nu/r)^4$.  Because the elastic scattering which populates the halo spreads neutrinos out to trajectories which have wider intersection angles while conserving their overall numbers, the $\nu-\nu$ contribution to Equation (2) which includes the halo is softer than $\propto (R_\nu/r)^4$.  This softening leads to a larger scale height for the $\nu-\nu$ potential and results in moderately more adiabatic conditions for collective neutrino oscillation.  While the exact degree of increase in adiabaticity is dependent on the details of the supernova model under consideration, the trend of increased adiabaticity when including halo neutrinos is uniform across all studied in this work.  Increasing the adiabaticity of collective neutrino oscillation leads to more complete and more step-like spectral swap features in the neutrino energy distribution.  An example of this can be seen clearly in the top panels of Figure~\ref{fig:spectra}, where the spectral swap between $\nu_{\rm e}-\nu_{\rm y}$ is greatly sharpened by the inclusion of the halo neutrinos in the collective neutrino oscillation calculation.

To quantify the comparison of received neutrino signals from the Z9.6 model, we have taken several limiting cases for fluxes of neutrinos generated by our flavor transformation calculations, one including matter effect only (no collective neutrino oscillation), one including radially emitted neutrinos only (no-halo), and radially emitted neutrinos including the outward directed halo neutrinos (with-halo), and used the SNOwGLoBES software package~\cite{snowglobes} to model the detected signal corresponding to several time snapshots of the explosion.  We have chosen to compare event rates for inverse beta-decay (IBD) in Super-Kamiokande (SK) (assuming the detector has completed doping with Gd, allowing for the tagging of IBD events), and $\nu_{\rm e} - \,{^{40}}{\rm Ar}$ capture in a $40\,\rm kt$ liquid argon (LAr) detector, projected to be DUNE.  Our reasoning behind considering the IBD rate in SK (SK) rather than the event rate in Hyper-Kamiokande (HK) is driven by our interest in the spectral distortions created by collective neutrino oscillation signals.  HK  will have a fiducial mass of $190\, \rm kt$ of water, compared to SK's fiducial mass of $22.5 \, \rm kt$, which will dramatically increase the total number of neutrino events and make HK considerably more sensitive to fast time variations of the SN neutrino burst.  However, because the HK detector will not be Gd-doped it will not be able to identify IBD events uniquely.  The events observed by HK will be a convolution of all neutrino detection channels (both capture and elastic scattering) for all flavors of neutrinos.  Because collective neutrino oscillation signals are typified by spectral swaps between neutrino flavors, a single flavor detection channel is superior in identifying the presence of collective neutrino oscillation, so we have elected to examine the SK IBD signal due to SK-Gd's ability to tag the $\bar\nu_{\rm e}$ with high efficiency.  For all analyses we conduct, event rates are calculated fixing the separation distance between Earth and the CCSN to $d = 10\,\rm kpc$.

\begin{figure*}[htbp]
    \begin{minipage}{0.5\hsize}
	\centering
    \includegraphics[width=1.0\linewidth]{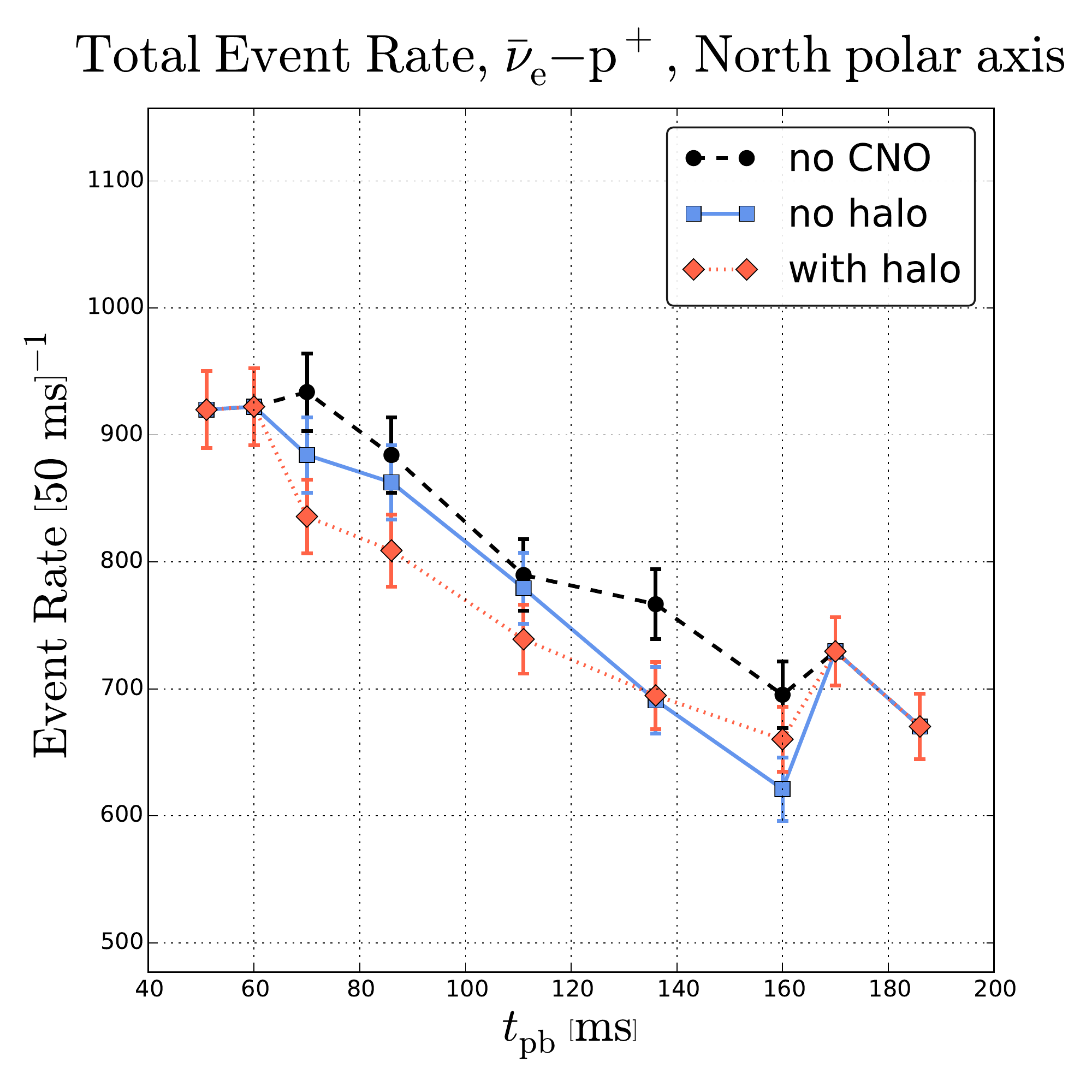}
    \end{minipage}	
    \begin{minipage}{0.5\hsize}
    	\centering
    \includegraphics[width=1.0\linewidth]{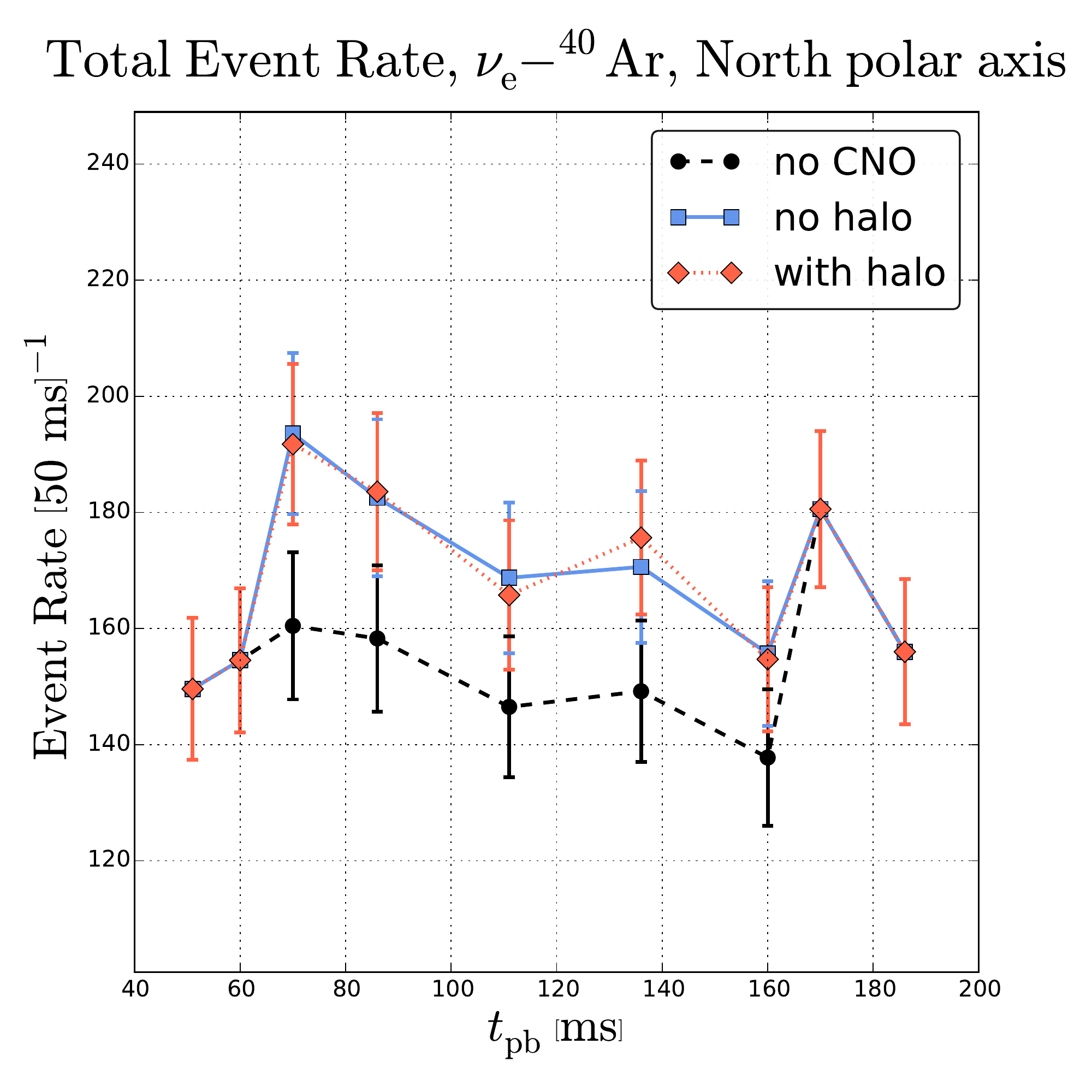}
        \end{minipage}
\caption{Left panel: the time evolution of the inverse beta decay event rate at SK.  Right panel: the time evolution of the electron neutrino-${^{40}}{\rm Ar}$ capture rate at DUNE.}
\label{fig:SK_event}
\end{figure*}

\begin{figure*}[htbp]
    \begin{minipage}{0.5\hsize}
	\centering
    \includegraphics[width=1.0\linewidth]{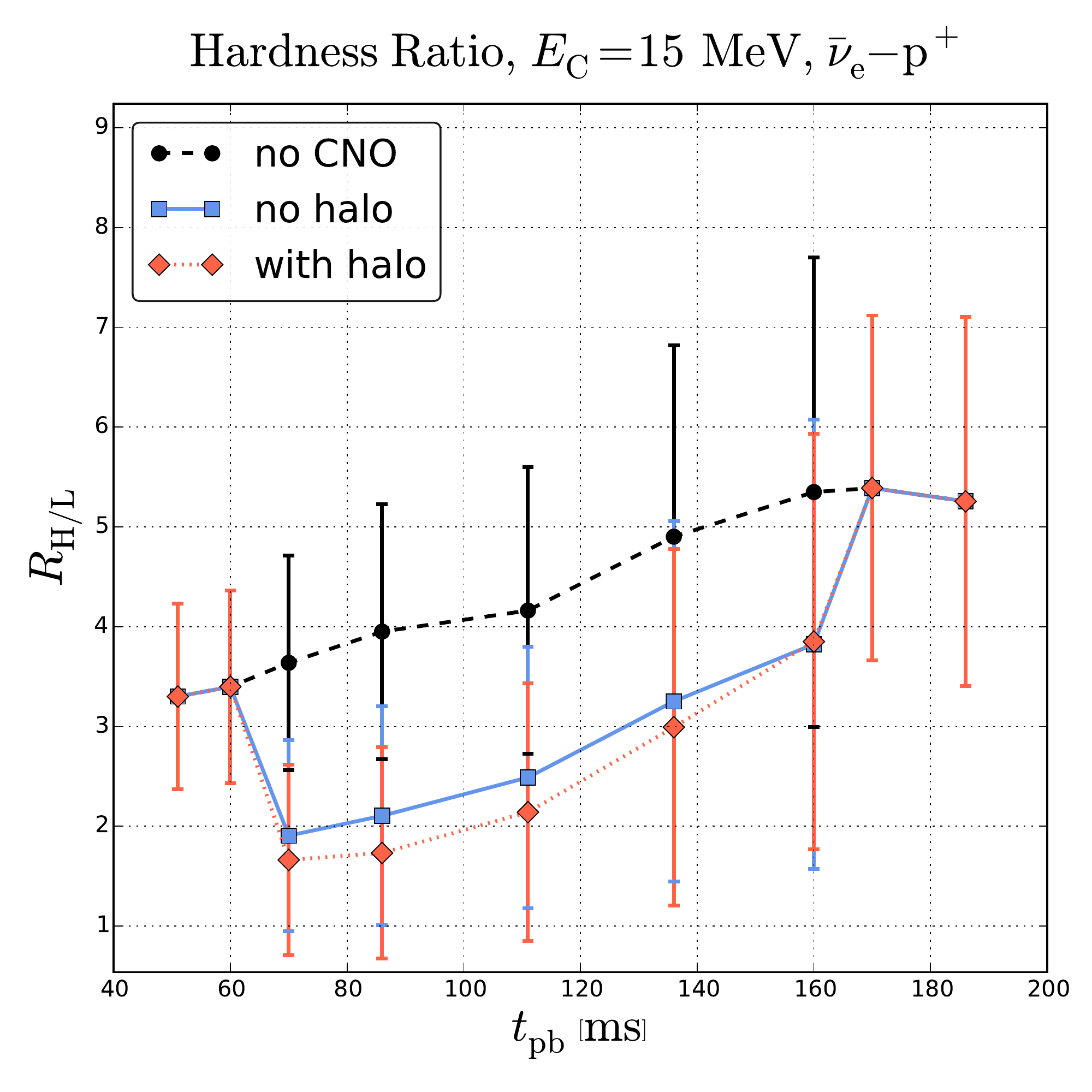}
    \end{minipage}
    \begin{minipage}{0.5\hsize}
    \centering
    \includegraphics[width=1.0\linewidth]{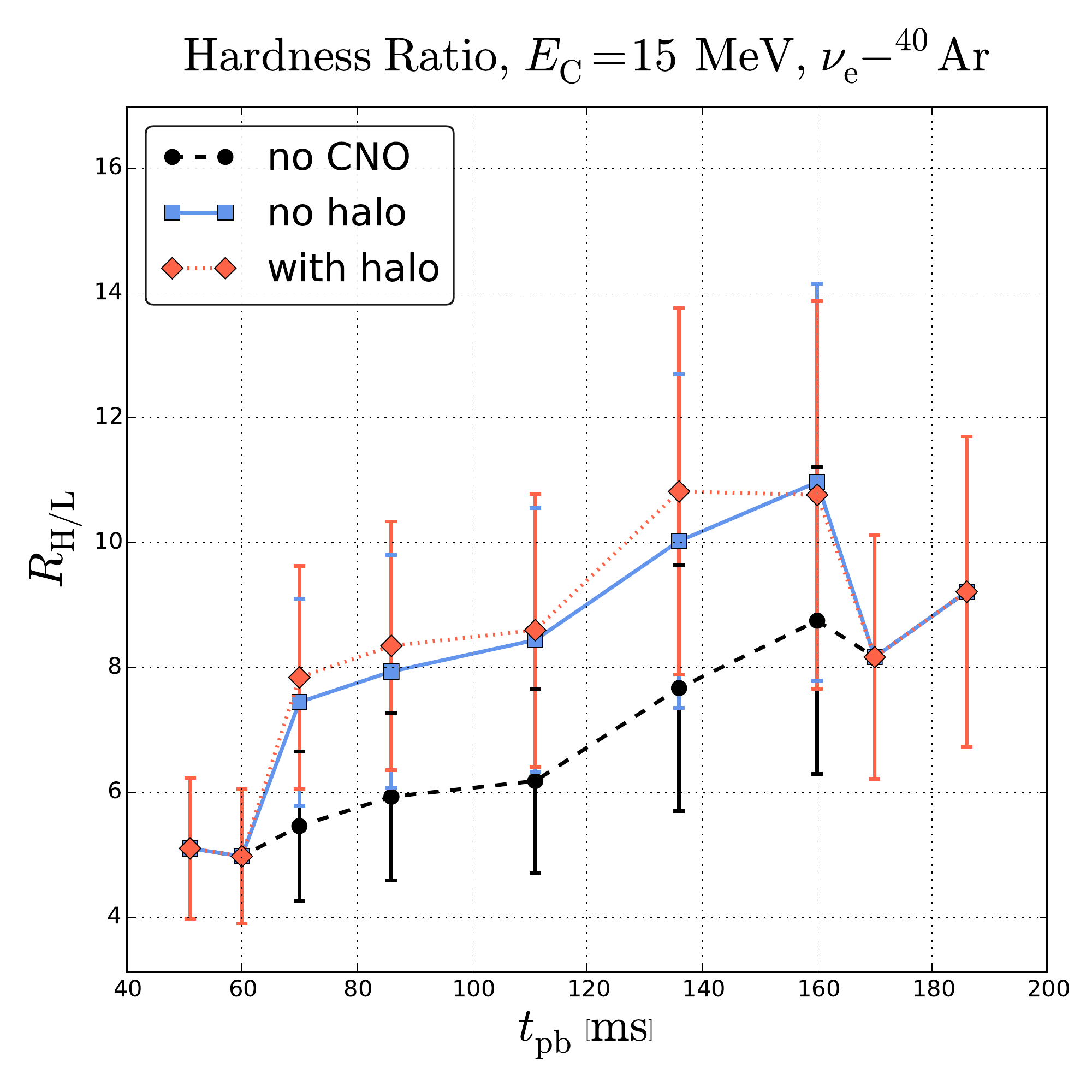}
    \end{minipage}
	\caption{Left panel: the time evolution of the Hardness ratio, $R_{\rm H/L}$, for the inverse beta decay event channel at SK.  Right panel: the time evolution of the Hardness ratio, $R_{\rm H/L}$, for the electron neutrino-${^{40}}{\rm Ar}$ capture channel at DUNE.}
	\label{fig:SK_Hardness}
\end{figure*}

Figure~\ref{fig:SK_event} shows the results for the total event rate for IBD in SK on the left panel and the total event rate for $\nu_{\rm e} - \,{^{40}}{\rm Ar}$ capture in DUNE on the right.  Note that because we chose to bin snapshots more closely near the onset and end of the collective neutrino oscillation epoch, the inter-snapshot spacing of data points in Figure~\ref{fig:SK_event} is not necessarily equal to the integration time used to calculate the event rate, which is $50\, \rm ms$.  From the total event rates alone, it is clear that the onset of collective neutrino oscillation for this CCSN model can be detected through the divergent trends in the event rates of both experiments.  When the Z9.6 simulation neutrino emission is modified by matter effects alone, the event rates for $\nu_{\rm e}$ and $\bar\nu_{\rm e}$ trend together after the passage of the neutronization burst.  However, there is a clear divergence in the relative event rates for $\nu_{\rm e}$ vs $\bar\nu_{\rm e}$ when collective neutrino oscillation is included, with a significant increase in $\nu_{\rm e}$ capture in LAr contemporaneous with a significant decrease in $\bar\nu_{\rm e}$ capture in inverse beta decay.  This relative shift in the event rates continues for the entirety of the collective neutrino oscillation epoch, before the total event rates in both experiments abruptly return to trending together at $t_{\rm pb} = 170\,\rm ms$.  The total event rate in LAr does not show any significant difference between the \lq no-halo\rq\ and \lq with-halo\rq\ cases, while the total events received during the collective neutrino oscillation epoch for SK are reduced an additional $10\%$ when including the halo in collective neutrino oscillation calculations.

Of course, the total event rate omits all of the information contained in the energy distribution of the observed signals for our proxy detectors.  We might also consider measures which are sensitive to the energy dependence of $\nu_{\rm e}$ and $\bar\nu_{\rm e}$ events.  To this end we introduce the \lq\lq Hardness Ratio\rq\rq , $R_{\rm H/L}$, which splits the event rates in each detection channel into two bins with detected neutrino energies above and below a cutoff energy, $E_{\rm C} = 15\,\rm MeV$.  This gives,
\begin{equation}
R_{\rm H/L} = \frac{N\vert_{E>E_{\rm C}}}{N\vert_{E<E_{\rm C}}}\, ,
\label{RHL}
\end{equation}
with a distinct $R_{\rm H/L}$ for each detection channel from which we consider.  Figure~\ref{fig:SK_Hardness} shows the Hardness ratio for inverse beta decay detection at SK on the left panel and the right panel shows the Hardness ratio for $\nu_{\rm e} - \,{^{40}}{\rm Ar}$ capture detection at DUNE.  

Much like the total event rates during this epoch of the CCSN explosion, the spectral hardness of $\nu_{\rm e}$ and $\bar\nu_{\rm e}$ are expected to trend together during this period of the Z9.6 CCSN.  $R_{\rm H/L}$ gradually increases for both detection channels as the proto-neutron star at the core of the explosion cools from neutrino emission, causing the spectra of all emitted neutrinos to stiffen.  With the onset of collective neutrino oscillation we see that again the behavior of simultaneously divergent trends, this time for in $R_{\rm H/L}$ in both detection channels.  However, the shifts in the Hardness ratio are less statistically significant than the collective neutrino oscillation induced shifts in the total event rate.  Likewise, there is very little distinction between the \lq no-halo\rq\ and \lq with-halo\rq\ cases for the $R_{\rm H/L}$ ratio.

\begin{figure*}[htbp]
    \begin{minipage}{0.5\hsize}
	\centering
    \includegraphics[width=1.0\linewidth]{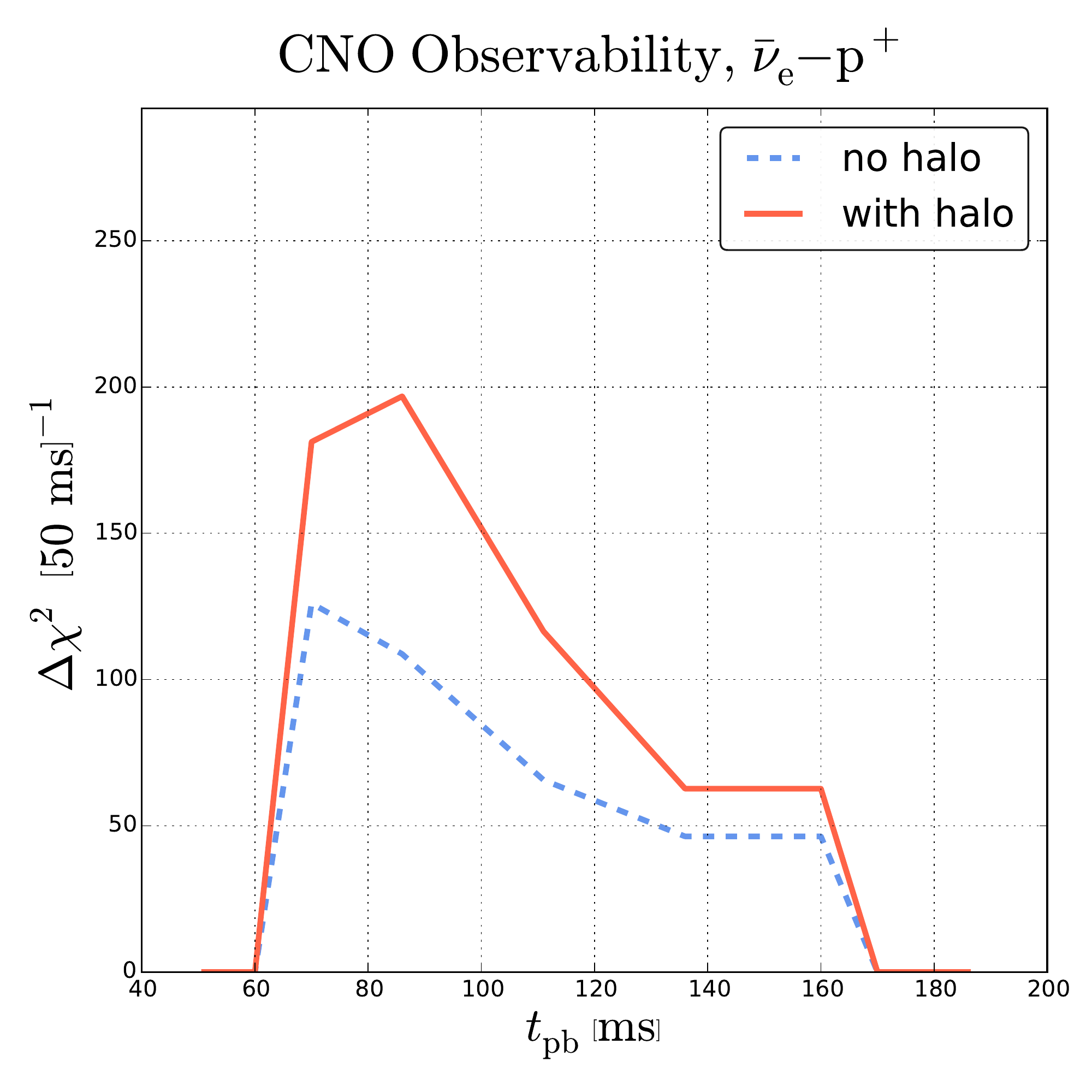}
    \end{minipage}
    \begin{minipage}{0.5\hsize}
         \centering
    \includegraphics[width=1.0\linewidth]{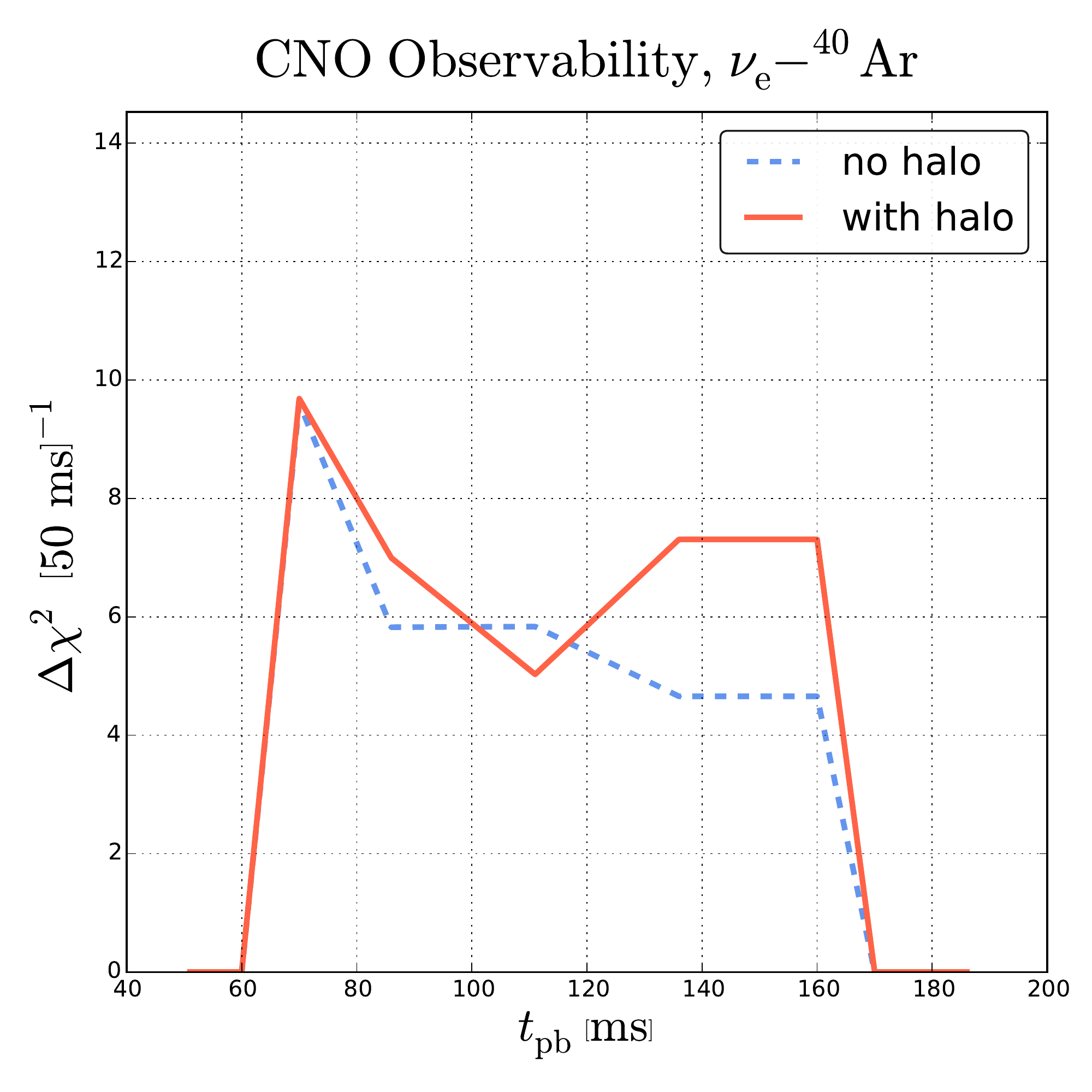}
    \end{minipage}
	\caption{Left panel: the time evolution of $\Delta \chi^2$ for the inverse beta decay event channel at SK.  Right panel: the time evolution of $\Delta \chi^2$, for the $\nu_{\rm e}-{^{40}}{\rm Ar}$ capture channel at DUNE.  For both cases we take the absence of collective neutrino oscillation to be the null hypothesis.}
	\label{fig:SK_deltachisq}
\end{figure*}

To restore as much of the shape information as possible to our predicted signals from the Z9.6 simulation, we have performed a basic $\Delta \chi^2$ hypothesis test on the received event distributions (assuming a uniform $\Delta E = 4\,\rm MeV$ for energy bins between $0-60\,\rm MeV$ and $50\,\rm ms$ integrated observing time), taking the \lq no collective neutrino oscillation\rq\ case as the null hypothesis for each channel.  This will give us a measure of the raw statistical potential to extract information from the collective neutrino oscillation epoch of the Z9.6 signal.    

Shown in the left panel of Figure~\ref{fig:SK_deltachisq} are the results for the inverse beta decay channel in SK and results for $\nu_{\rm e} - \,{^{40}}{\rm Ar}$ capture detection at DUNE are shown in the right panel.  These two figures illustrate the importance of including the halo neutrinos in collective neutrino oscillation calculations.  Typically, although not uniformly, collective neutrino oscillation signals are more easily discriminated from non-collective neutrino oscillation signals when halo neutrinos are accounted for in both SK and DUNE.  For the Z9.6 CCSN simulation, the effect is much more pronounced in SK, where there is a $\sim 60\%$ increase in $\Delta \chi^2$ integrated over the collective neutrino oscillation epoch.  Although we do not attempt to make a detailed spectral reconstruction from the received signals in SK and DUNE, both panels of Figure~\ref{fig:SK_deltachisq} taken together indicate that the spectral shape of the collective neutrino oscillation signal is more easily discriminated from thermal neutrino emission when halo neutrinos are included in collective neutrino oscillation calculation.  Put another way, because previous studies have omitted halo neutrinos in their collective neutrino oscillation calculations, their predictions for extracting meaningful signals from a supernova neutrino burst are likely to have been {\it overly pessimistic} with regard to strength of collective neutrino oscillation signals.

\section{Conclusions}
\label{sec:4}

We have made the first multi-angle calculation of neutrino flavor evolution in the iron-core collapse supernova environment which includes the population of neutrinos scattered into the wide angle halo.  We have shown that there are qualitative and quantitative consequences for resulting neutrino oscillation signatures, relative to collective neutrino oscillation calculations which omit the halo neutrinos.  We have shown that these changes have implications for the detectability of collective neutrino oscillation in the CCSN neutrino burst signal of a $9.6 M_\odot$ progenitor.

The physical circumstances which proceed from the evolution of the explosion of the Z9.6 progenitor star produce an environment which is conducive to calculating collective neutrino oscillation with the inclusion of halo neutrinos.  Multi-angle suppression of neutrino flavor transformation deep within the envelope is present both prior to $t_{\rm pb} = 70\,\rm ms$ and after $t_{\rm pb} = 170\,\rm ms$. While collective neutrino oscillation is unsuppressed for the intervening $100\, \rm ms$, this intermediate window coincides with the epoch when the explosion shock front has not yet grown outward to the initial radius of collective neutrino flavor conversion. In other words, the relatively low radius of the shock creates a neutrino halo which is predominantly outward directed as far as collective neutrino oscillation is concerned. By the epoch of collective neutrino oscillation, inward directed neutrino flux is found to be a sub-leading order contribution to the flavor changing Hamiltonian. This enables the multi-angle calculation of collective neutrino oscillation for the outgoing flux of neutrinos during this window of time.

We compare the results of our collective neutrino oscillation calculations with those which omit the halo neutrino population, as well as those which omit collective neutrino oscillation entirely, and find important results.  Firstly, we find that the onset of the collective neutrino oscillation epoch in the neutrino burst signal is clearly distinguishable from thermal emission.  We also find that redistribution of neutrinos to wider emission trajectories via coherently enhanced neutral current nucleus scattering produces collective neutrino oscillation evolution which is more adiabatic than one would predict when omitting the halo population.  This changes the development of the spectral swap features which are hallmarks of collective neutrino oscillation signals, reducing the effects of collective flavor oscillation decoherence, shifting spectral swap energies, and increasing the sharpness of swap transitions.  Importantly, we find that detected collective neutrino oscillation signals tend to be more clearly distinguished from thermal emission when including the halo neutrinos.

Looking forward we are hopeful for the detection of collective neutrino oscillation signals from terrestrial neutrino detectors in the event of a galactic CCSN.  Our results show that the neutrino halo, which is a generic phenomenon common to all CCSNe, tends to enhance the non-thermal features of collective neutrino oscillation signals, potentially rendering them more easily detectable.  Conversely, this raises the possibility that previous studies which omit the presence of halo neutrinos in collective neutrino oscillation may be overly pessimistic in their calculations of the detectability of collective neutrino oscillation signals.  While our study here is a narrow sampling of the potential variety of neutrino burst signals from CCSNe, it suggests that the observational opportunities for studying CCSNe through neutrino messengers is richer than previously imagined.

\begin{acknowledgments}
This work has been partly supported by Grant-in-Aid for Scientific Research 
(JP17H01130, 
JP17K14306, 
JP18H01212) 
from the Japan Society for Promotion  of Science (JSPS) and the Ministry of Education, Science and Culture of Japan (MEXT, Nos. 
JP17H05206, 
JP17H06357, 
JP17H06364, 
JP24103001), and by the Central Research Institute of Stellar Explosive Phenomena (REISEP) at Fukuoka University and the associated projects (Nos.\ 171042,177103),
and JICFuS as a priority issue to be tackled by using Post `K' Computer. 
S.H.\ is supported by the U.S.\ Department of Energy under Award No.\ DE-SC0018327 and NSF Grants Nos.\ AST-1908960 and PHY-1914409. 
Numerical computations were in part carried out on Cray XC50 at Center for Computational Astrophysics, National Astronomical Observatory of Japan.
\end{acknowledgments}

\appendix
\section{Binning Halo Neutrino Angular Distributions}
Principally, we are interested in deriving an algorithm for binning the wide angle scattered neutrino population {in environments which may cause the solution of neutrino flavor evolution to be susceptible to numerical error, typically from rapid onset of collective neutrino oscillation. This is a necessity for robust collective neutrino oscillation calculations}. We have found that although it is relatively straightforward to calculate the initial distribution of halo neutrinos, the angular grid which is efficient for numerically converged collective neutrino oscillation calculations is non-trivially related to the post-hoc results of the halo calculation itself.  As a result, we have found it necessary to perform a bespoke calculation of the angular grid spacing on which to initialize the neutrino density matrices for the collective neutrino oscillation code.  The halo calculation computes the density/intensity of neutrino radiation emerging from the surface of the halosphere, with radius $R_{H}$, which is the spherical surface at which we have deemed the incoming halo neutrino flux to be negligible.  This gives an initial condition for emission in terms of neutrino energy $E_\nu$, the emission angle $\vartheta$, and flavor state, $\alpha$.  An example of this initial configuration can be seen in Figure~\ref{fig:imp_intensity}, which shows the neutrino emission intensity as a function of the impact parameter, $b = R_{H}\sin\vartheta$.

The neutrino intensity is normalized such that,
\begin{equation}
\int_{0}^{1}\int_{0}^{\infty} \frac{I_{\nu_\alpha}\left(E_{\nu_\alpha},\vartheta \right)}{4 \pi R_{H}^2} \frac{\left( R_{H}/r\right)^2 \cos\vartheta}{\sqrt{1 - \left( R_{H}/r\right)^2 \sin^2 \vartheta}} d\cos\vartheta\, d E_{\nu_\alpha} = \frac{1}{2}\frac{L_{\nu_\alpha}}{4\pi r^2 \langle E_{\nu_\alpha} \rangle} = \rho_{\nu_\alpha}\, ,
 \end{equation}
 where $r$ is the radius where the collective neutrino oscillation calculation is to be started, $\alpha$ is the neutrino flavor index,  $L_{\nu_\alpha}$ is the neutrino luminosity, and $\langle E_{\nu_\alpha} \rangle_\alpha$ is the average neutrino energy, and $\rho_{\nu_\alpha}$ is the number density of neutrinos contributing to the collective neutrino oscillation calculation.
 
 Because we are interested in recovering a binning scheme for the angular distribution, we will preemptively perform the energy integration and discuss the number density contribution for each neutrino species as a function of emission angle alone,
 \begin{equation}
 \rho_{\nu_\alpha}\left(\vartheta\right) =  \frac{I_{\nu_\alpha}\left(\vartheta \right)}{4 \pi R_{H}^2} \frac{\left( R_{H}/r\right)^2 \cos\vartheta}{\sqrt{1 - \left( R_{H}/r\right)^2 \sin^2 \vartheta}}\, .
 \end{equation}
From this perspective, we are in search of a set of angels, $\vartheta_j$, and bin widths, $\Delta\cos\vartheta_j$, such that we numerically recover the definite integral,
\begin{equation}
\int^1_0 \rho_{\nu_\alpha}\left(\vartheta\right) d\cos\vartheta = \rho_{\nu_\alpha} \approx \sum_{j=1,N_\vartheta} \rho_{\nu_\alpha}\left(\vartheta_j\right) \Delta\cos\vartheta_j\, ,
\end{equation}
while simultaneously minimizing the total number of angular bins, $N_\vartheta$, needed to achieve numerically converged collective neutrino oscillation results.

\section{Selecting the Weight Function} 
We will work from the perspective that the most efficient binning scheme is one which places bins most densely where the flavor evolution is most sensitive to changes in the neutrino flavor states.  To this end we must posit a weight function which assigns a relative importance to a given choice of $( \vartheta_j,\Delta\cos\vartheta_j )$, and choose the set of all $\{ (\vartheta_j,\Delta\cos\vartheta_j) \}$ such that each pair of the set has equal weight.  The simplest possible guess is to weight the importance of bins by their relative contribution to the coherent forward scattering Hamiltonian,
\begin{eqnarray}
W^{test}_{ij} &&= \int_{\cos\vartheta_i}^{\cos\vartheta_f} (1-\cos\theta_{ij})\sum_{\alpha} \left[\rho_{\nu_\alpha} \left(\vartheta\right) - \rho_{\bar\nu_\alpha} \left(\vartheta\right) \right] d\cos\vartheta \notag \\
&&= \frac{1}{N_\vartheta}\int_0^1  (1-\cos\theta_{ij})\sum_{\alpha} \left[\rho_{\nu_\alpha} \left(\vartheta\right) - \rho_{\bar\nu_\alpha} \left(\vartheta\right) \right] d\cos\vartheta \, ,
\end{eqnarray}
where $\cos\theta_{ij}$ is the intersection angle of trajectory j with the reference trajectory, i, with,
\begin{eqnarray}
\vartheta_j &&= \cos^{-1}\left(\frac{\cos\vartheta_f - \cos\vartheta_i}{2}\right)\\ 
{\rm and} \, \Delta\cos\vartheta_j &&= \cos\vartheta_f - \cos\vartheta_i \, .
\end{eqnarray}
However, this particular choice of the weight function has potential problems.  Specifically, because of the $\propto E_\nu^2$ cross section of the zero energy transfer direction changing scattering which populates the Halo, the number density of each species scattered along a trajectory varies with the second moment of the spectral energy distribution of that species.  Because of this, the weight function defined above is guaranteed to have zeros if,
\begin{equation}
Sign\sum_\alpha\left[\frac{L_{\nu_\alpha}}{\langle E_{\nu_\alpha}\rangle} - \frac{L_{\bar\nu_\alpha}}{\langle E_{\bar\nu_\alpha}\rangle} \right]  \neq Sign\sum_\alpha\left[\frac{L_{\nu_\alpha}\langle E^2_{\nu_\alpha}\rangle}{\langle E_{\nu_\alpha}\rangle} - \frac{L_{\bar\nu_\alpha}\langle E^2_{\bar\nu_\alpha}\rangle}{\langle E{\bar\nu_\alpha}\rangle} \right] \, ,
\end{equation} 
where the left hand side is proportional to number densities in the bulb neutrino emission and the right hand side is proportional to number densities in the halo emission.  A trajectory which has zero weight will result in an anomalously large bin.  However, because collective neutrino oscillation does not conserve the quantity $\rho_{\nu_\alpha} \left(\vartheta\right)-\rho_{\bar\nu_\alpha} \left(\vartheta\right)$, a bin which has an initial configuration where $\rho_{\nu_\alpha} \left(\vartheta\right)-\rho_{\bar\nu_\alpha} \left(\vartheta\right) \approx 0$ is not guaranteed to remain that way.  As a result, the above weight function may be in danger of producing a binning scheme which is good for the initial neutrino flavor distribution, but potentially unbalanced after collective neutrino oscillation has begun.

An alternative weight scheme is to simply sum over the total number density of neutrinos and anti-neutrinos along each trajectory.  This gives the weight function,
\begin{eqnarray}
W_{ij} &&= \int_{\cos\vartheta_i}^{\cos\vartheta_f} (1-\cos\theta_{ij})\sum_{\alpha} \left[\rho_{\nu_\alpha} \left(\vartheta\right) + \rho_{\bar\nu_\alpha} \left(\vartheta\right) \right] d\cos\vartheta \notag \\
&&= \frac{1}{N_\vartheta}\int_0^1  (1-\cos\theta_{ij})\sum_{\alpha} \left[\rho_{\nu_\alpha} \left(\vartheta\right) + \rho_{\bar\nu_\alpha} \left(\vartheta\right) \right] d\cos\vartheta \, ,
\label{finalweight}
\end{eqnarray}
again with $\vartheta_j$ and $\Delta\cos\vartheta_j$ as defined in Equations (B2) and (B3).  This choice of weight function does not have any zeros which are produced by the change in the spectral energy distribution of neutrinos as a function of the angle $\vartheta$.  While the binning configuration is not guaranteed to remain optimal throughout the later collective neutrino oscillation evolution, weighting evenly in the total number density approximately distributes weight by the {\it maximum} potential contribution from each bin in the event of total flavor conversion, as opposed to the potential contribution of the initial flavor configuration.

\section{Selecting the Reference Trajectory} 
\begin{figure}[h]
	\centering
	\includegraphics[angle=0,width=0.6\textwidth]{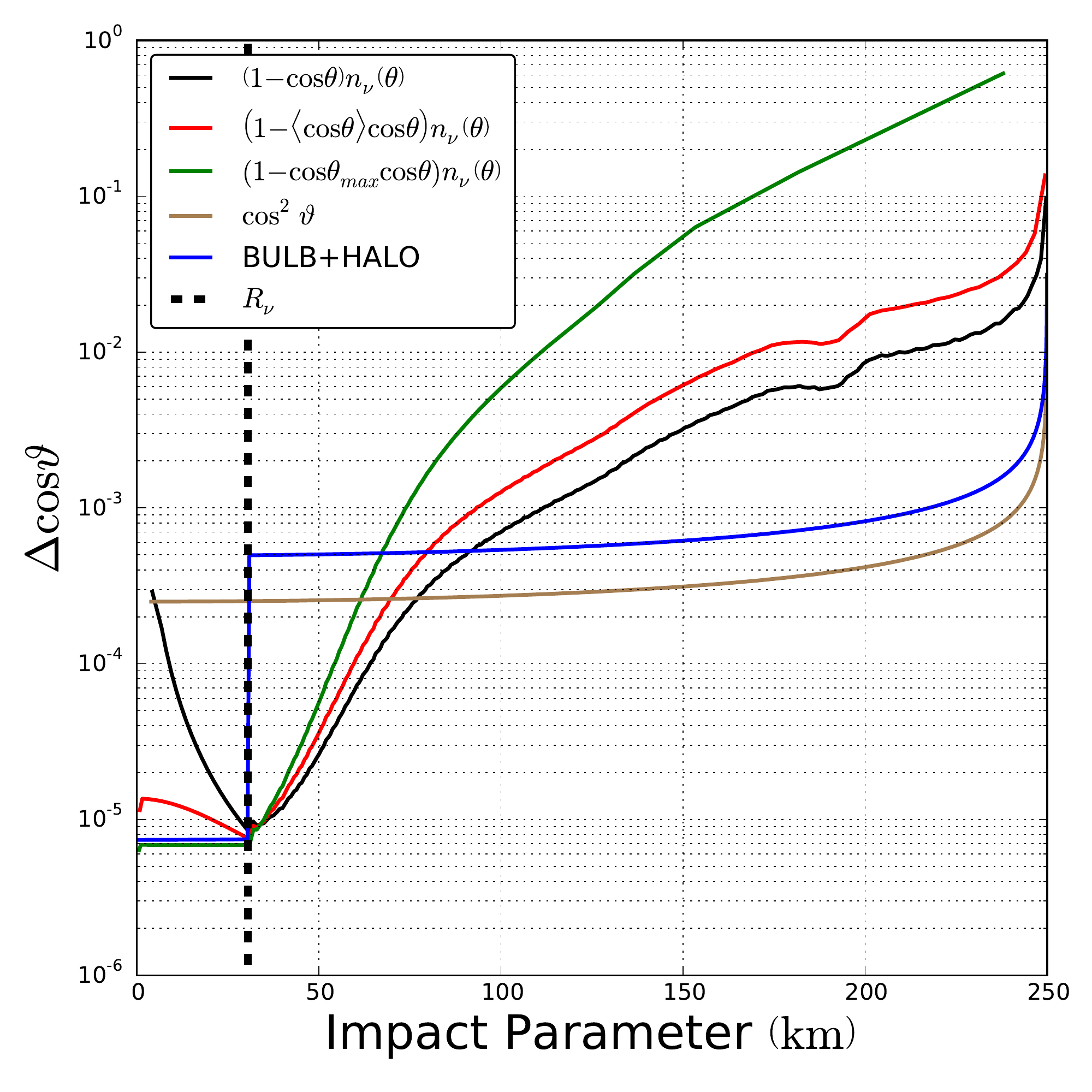}
	\caption{A comparison of various schemes for binning the total neutrino emission, showing the bin size $		\Delta\cos\vartheta_j$ as a function of the impact parameter of the trajectory, $b_j = R_{H}\sin\vartheta_j$.  The dashed black line denotes the location of the neutrino sphere near the surface of the proto-neutron star.}
\label{fig:bin_scheme}
\end{figure}

Because the Hamiltonian contribution of the neutrinos is cast in terms of the intersection angle of two trajectories, $\cos\theta_{ij} = \cos\theta_{i}\cos\theta_{j} $, and we are specifying the binning scheme for all trajectories $\{ (\vartheta_j,\Delta\cos\vartheta_j) \}$, any choice we care to make for trajectory $i$ will alter the resulting $W_{ij}$ along with the set of solutions $\{ (\vartheta_j,\Delta\cos\vartheta_j) \}$.  The intersection angle of trajectory, $\theta_i$, is given in terms of the emission angle, $\vartheta_i$,
\begin{equation}
\cos\theta_i = \sqrt{1 - (R_{H}/r)^2\sin^2\vartheta_i}\, ,
\end{equation}
which gives two limiting cases.  The first limit is to employ the radial trajectory as a reference in computing $W_{ij}$, taking $\cos \theta_i = 1$.  The second limit is to employ the most tangential emission trajectory as the reference for computing $W_{ij}$, taking $\cos\theta_{i}= \cos\theta_{max} = \sqrt{1 - (R_{H}/r)^2}$.  

In Figure~\ref{fig:bin_scheme} we compare a variety of different potential angular binning schemes, $\{ (\vartheta_j,\Delta\cos\vartheta_j) \}$ using the neutrino emission intensity shown in Figure~\ref{fig:imp_intensity}, in terms of the bin width.  The tan line, $\cos^2\vartheta$, is the historic choice for the BULB model and is optimal for uniform emission from the surface of the neutrino sphere, setting $\Delta\cos\vartheta \propto \cos^2\vartheta$.   The blue line, labeled BULB+HALO, is the ad hoc binning scheme used in Ref.~\cite{Cherry:2013a}, which was used in a context where collective neutrino oscillation numerical stability was much less sensitive to the choice of angular binning.  The black line shows the $\cos \theta_i = 1$ reference case for our proposed method, which generates relatively narrow bins in the inner-Halo region, $100\,\rm km > b > R_{\nu}$, but generates bins in the neutrino sphere emission region, $b < R_{\nu}$, which are an order of magnitude wider than the previous successful calculation for the  BULB+HALO binning scheme.  The green line shows the tangential emission reference case for $\cos\theta_{i}= \cos\theta_{max} = \sqrt{1 - (R_{H}/r)^2}$.  The $\cos\theta_{max}$ reference case produces very narrow bins in the neutrino sphere emission region but produces bins which are phenomenally wide in the outer regions of the Halo emission.  For example, the final bin has $\Delta\cos\vartheta_j$ which corresponds to $\Delta b = 100\,\rm km$.  This limit is unlikely to sufficiently resolve the angular flavor evolution in the Halo population.

As a compromise between these two limits, we use the neutrino number density weighted average intersection angle.  We define the average reference trajectory, $\langle \cos\theta\rangle$,
\begin{equation}
\langle \cos\theta\rangle = \frac{\int_0^1 \sqrt{1 - (R_{H}/r)^2\sin^2\vartheta}\times\sum_{\alpha} \left(\rho_{\nu_\alpha} \left(\vartheta\right) + \rho_{\bar\nu_\alpha} \left(\vartheta\right) \right) d\cos\vartheta}{\int_0^1 \sum_{\alpha} \left(\rho_{\nu_\alpha} \left(\vartheta\right) + \rho_{\bar\nu_\alpha} \left(\vartheta\right) \right) d\cos\vartheta} \, .
\end{equation}
Using the average intersection angle as the reference trajectory, $\cos\theta_i = \langle \cos\theta\rangle$, produces the red line in Figure~\ref{fig:bin_scheme}.  We can see that the $\langle \cos\theta\rangle$ reference trajectory produces relatively narrow bins in the neutrino sphere emission region as well as relatively narrow bins in the Halo region.  While average intersection angle reference never produces binning solution which are as narrow as the limiting case solutions, it is always within a factor of two of the narrowest binning limiting case solution.  Using the $\cos\theta_i = \langle \cos\theta\rangle$ reference trajectory in conjunction with the weight scheme described by Equation~\ref{finalweight} to select the angular bin distribution is typically sufficient to produce numerically converged results with $N_\vartheta \sim 1000\, -\, 2000$ angular bins.

\section{$15 M_\odot$ Progenitor}

\begin{figure*}[htbp]
	\begin{minipage}{0.5\hsize}
	    \centering
        \includegraphics[width=1.0\linewidth]{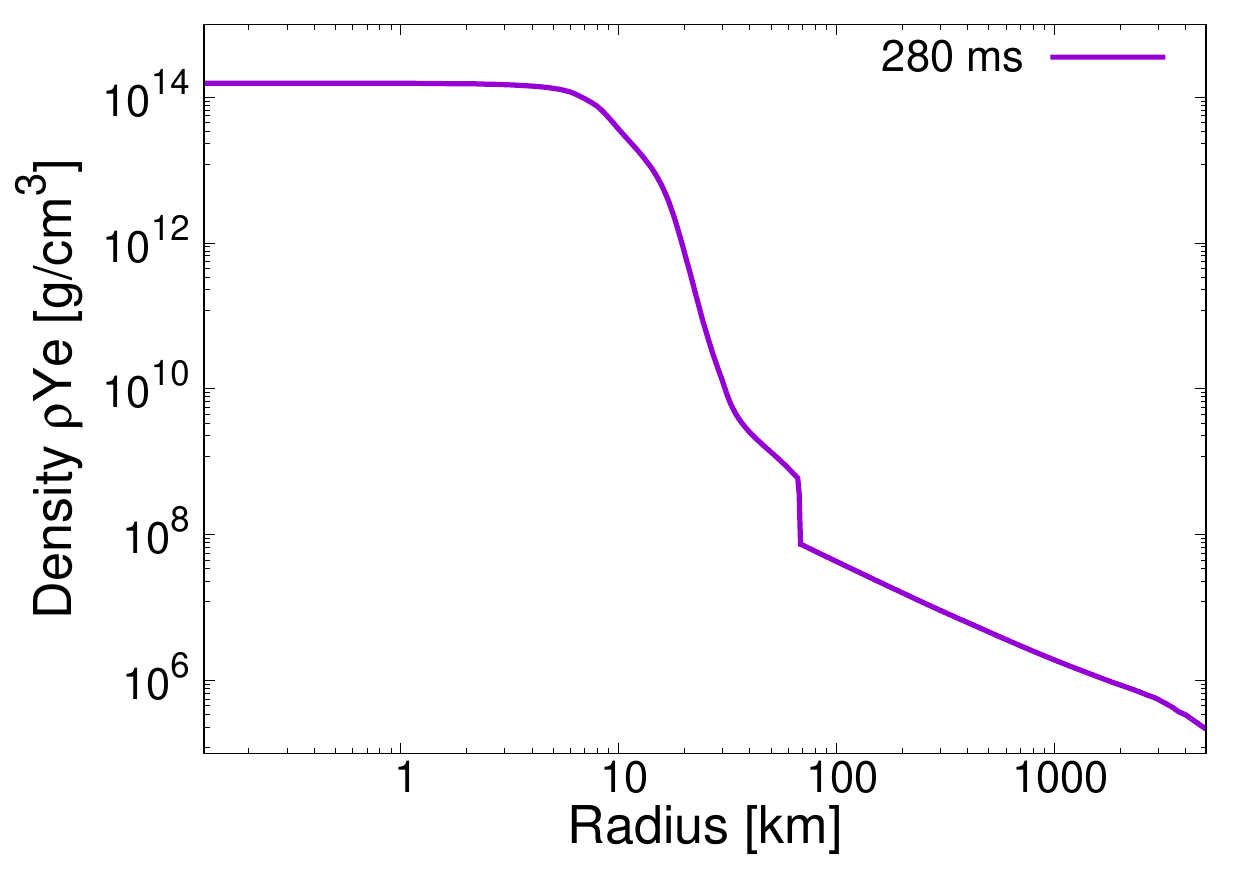}
	\end{minipage}
	\begin{minipage}{0.5\hsize}
	    \centering
	    \includegraphics[width=1.0\linewidth]{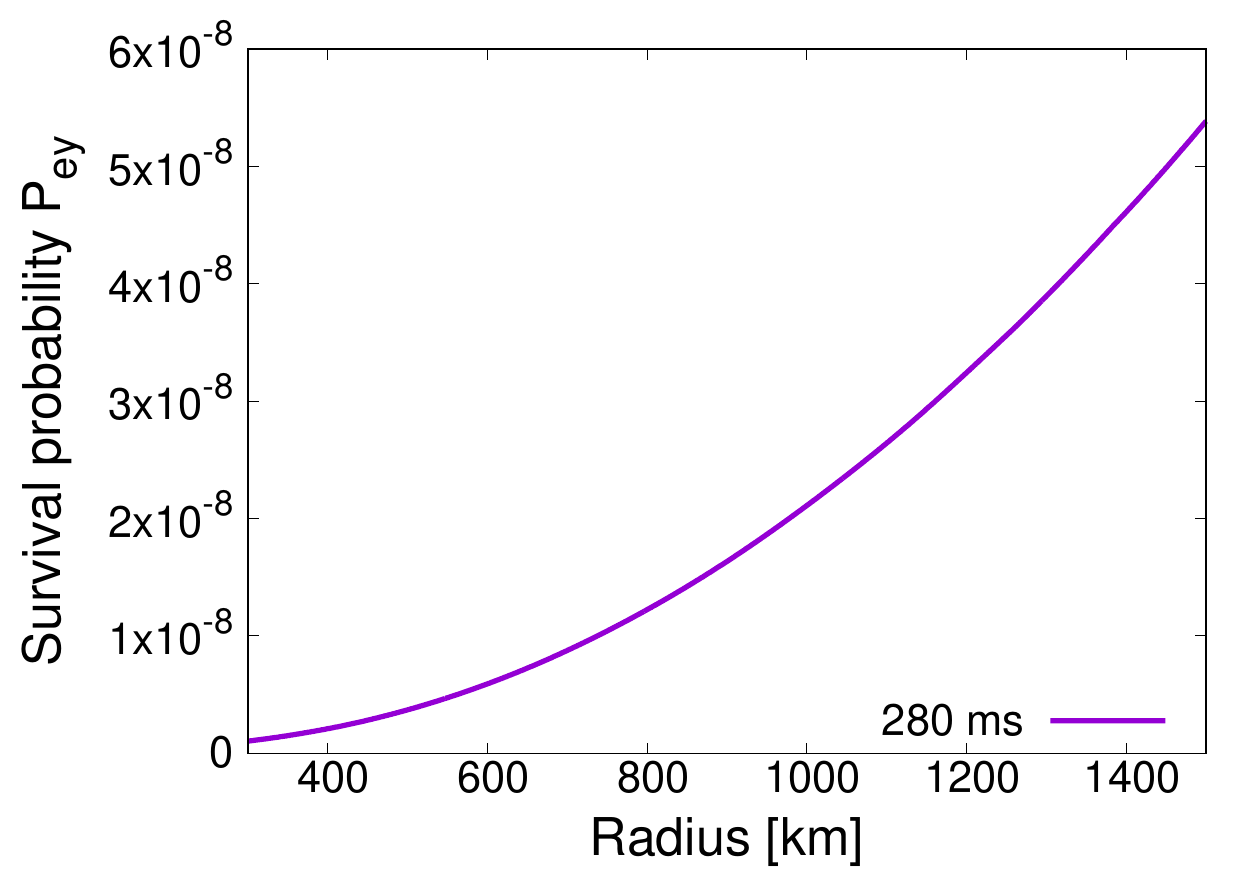}
	\end{minipage}	\caption{
	Left: density profile of $15 M_{\odot}$ progenitor at $280\mathrm{~ms}$. Right: The radial evolution of the survival probability from electron neutrino to non-electron type.
	}
	\label{fig:s15_case}
\end{figure*}

In order to provide comparison to previous work \cite{Sarikas:2012b}, we will briefly discuss our results for neutrino flavor transformation for a more massive progenitor case, the 1D simulation of the $15 M_{\odot}$ progenitor of Woosley and Weaver \cite{Woosley_1995} used in the work of Sarikas et al.~\cite{Sarikas:2012b}.
The left panel in Figure \ref{fig:s15_case} shows the density profile of the $15 M_{\odot}$ progenitor at post-bounce time, $t_{\mathrm{pb}}=280\mathrm{~ms}$.
Compared with our Z9.6 model in Figure \ref{fig:north_density}, the more massive progenitor exhibits a delayed shock revival and decreased shock radius of $70\mathrm{~km}$ at this time. The right panel in Figure \ref{fig:s15_case} shows the radial evolution of the survival probability of electron neutrinos to non-electron type, including the wide angle scattering of halo neutrinos.
Collective neutrino flavor transformation is completely suppressed and the small amplitude exponential growth is consistent with MSW effect driven flavor conversion.
The complete suppression of the self-induced conversion during the accretion phase is consistent with the linearized stability analysis predictions of \cite{Sarikas:2012b}.
This feature is common in CCSNe with dense envelopes like that of the $15 M_{\odot}$ star shown in Figure \ref{fig:s15_case}.


\providecommand{\href}[2]{#2}\begingroup\raggedright\endgroup

\end{document}